\documentclass[article,nojss,shortnames]{jss}

\usepackage{thumbpdf}
\usepackage{amsfonts,amstext,amsmath,amssymb,amsthm}
\usepackage{color}
\usepackage{rotating}
\usepackage{verbatim}
\usepackage{caption}
\usepackage{subcaption}

\usepackage{color}
\definecolor{sidebox_FIXME}{rgb}{1,1,0.2}

\author{Heidi Seibold \\ Universit\"at Z\"urich \\
   \And Achim Zeileis \\ Universit\"at Innsbruck \\
   \And Torsten Hothorn \\ Universit\"at Z\"urich}
\Plainauthor{Seibold, Zeileis, Hothorn}

\title{Individual Treatment Effect Prediction \\ for ALS Patients}
\Plaintitle{Individual Treatment Effect Prediction for ALS Patients}
\Shorttitle{Individual Treatment Effect Prediction for ALS Patients}

\Abstract{
A treatment for a complicated disease may be helpful for some but not all
patients, which makes predicting the treatment effect for new patients
important yet challenging.  Here we develop a method for predicting the
treatment effect based on patient characteristics and use it for predicting the
effect of the only drug (Riluzole) approved for treating Amyotrophic Lateral
Sclerosis (ALS).  Our proposed method of model-based random forests detects
similarities in the treatment effect among patients and on this basis computes
personalised models for new patients. The entire procedure focuses on a base
model, which usually contains the treatment indicator as a single covariate and
takes the survival time or a health or treatment success measurement as primary
outcome.  This base model is used both to grow the model-based trees within the
forest, in which the patient characteristics that interact with the treatment
are split variables, and to compute the personalised models, in which the
similarity measurements enter as weights. We applied the personalised models
using data from several clinical trials for ALS from the PRO-ACT database.  Our
results indicate that some ALS patients benefit more from the drug Riluzole
than others.  Our method allows shifting from stratified medicine to
personalised medicine and can also be used in assessing the treatment effect for
other diseases studied in a clinical trial.

}

\Plainkeywords{Personalised medicine, random forest, treatment effect, model-based recursive partitioning}

\Address{
  Heidi Seibold\\
  Institut f\"ur Epidemiologie, Biostatistik und Pr\"avention, Abteilung Biostatistik \\
  Universit\"at Z\"urich \\
  Hirschengraben 84, CH-8001 Z\"urich, Switzerland \\

  Achim Zeileis \\   
  Department of Statistics \\
  Faculty of Economics and Statistics \\
  University of Innsbruck \\
  Universit\"atsstr. 15 \\
  A-6020 Innsbruck, Austria \\

  Torsten Hothorn\\
  Institut f\"ur Epidemiologie, Biostatistik und Pr\"avention, Abteilung Biostatistik \\
  Universit\"at Z\"urich \\
  Hirschengraben 84, CH-8001 Z\"urich, Switzerland \\
  E-mail: \email{Torsten.Hothorn@R-project.org}\\
  URL: \url{http://user.math.uzh.ch/hothorn/} \\

}

\begin{document}
%% \footnote{\texttt{ctm.pdf} compiled \today~from \texttt{ctm.tex} version SVNREV}

\section{Introduction}

Amyotrophic lateral sclerosis (ALS) is a deadly disease that affects motor
neurons in the brain and spinal cord, i.e.\ the neurons responsible for
voluntary muscle control. Riluzole (Rilutek) is the only approved drug for this
disease to date. According to the \cite{ema_riluzole_2012}, Riluzole prolongs
the median survival of ALS patients, depending on the dose, by a few months.
Several side effects, such as sickness, weakness or increased liver enzyme
levels are mentioned \citep{ema_riluzole_2012}. Knowledge how Riluzole works on
the nervous system of ALS patients is limited. The PRO-ACT database
\citep{atassi_pro-act_2014} is the largest database containing clinical trial
data of ALS patients available and was initiated to retrieve more information
on the disease. It contains data from 17 ALS studies conducted between 1990 and
2010. Using these data, we aimed at finding out more about the effect of
Riluzole on the health and survival of patients.

Before statistical analysis and $p$-values entered into medical progress,
doctors treated patients individually based on their experiences and knowledge
\citep{weisberg_what_2015}.  Since the beginning of the ``golden age of
randomised clinical trials'', however, medication became more and more
standardised. Nowadays, much knowledge about the effect of drugs has accumulated,
cornerstone drugs such as antibiotics have been used for decades and many
diseases can be treated successfully; however, providing new drugs for the
general public is more difficult. Diseases such as ALS are too complex to treat
all patients in the same way. Therefore, there is a need to return to more
individualised treatments, but this time with the use of statistical concepts.

In the past years, there has been an immense effort towards personalised
medicine in the analysis of randomised controlled trials. The goal is to
identify predictive factors, i.e.\ factors that interact with the treatment
\citep{italiano_prognostic_2011}, such as biomarkers, other treatments, and
environmental circumstances. In the following, we will refer to these factors
as patient characteristics. Prognostic factors, i.e.\ factors that directly
affect the patient's outcome, are only of secondary interest, but should not be
neglected, because they not only change the general level of the outcome --
showing in the individual intercept -- but might also be predictive and
prognostic \citep{seibold_model-based_2015}.  For drugs for which the
biological mode of action is unknown,  predictive and prognostic factors should
first be identified in a data-driven way. New hypotheses can then be generated
and new trials can be planned based on these hypotheses.  In this first step we
are asking  \textit{whether} a certain patient characteristic is relevant and
not \textit{why}. 

Many new statistical methods in the field of stratified medicine, i.e.\
subgroup analysis, have been developed. Subgroup analyses aim at finding groups
of patients that have differential treatment effects.  Most of the methods are
based on recursive partitioning (trees) and/or interaction models
\citep{ciampi_tree-structured_1995, kehl_responder_2006,
dusseldorp_qualitative_2013, loh_regression_2014, tian_simple_2014,
foster_simple_2015, zhang_robust_2012}. The tree methods for subgroup analyses
have specialised splitting procedures for partitioning the patients into groups
with higher and lower treatment effect. Interaction models evaluate the
interaction between the treatment and given patient characteristics.  The idea
behind methods of subgroup analyses in general is to obtain a treatment effect
$\beta(\mathbf{z})$ that depends on the patient characteristics $\mathbf{z}$.
For example, the treatment effect could depend on the age of patients, in which
patients less than 40 years of age improve through the treatment, patients
between 40 and 60 do not improve and patients older than 60 years improve, but
less than the patients under 40 years:
\begin{align}
	\beta(\mathbf{z}) &= \left\{ \begin{array}{rl}
		1 & \text{if } z_\text{age} < 40 \\
                0 & \text{if } 40 \leq z_\text{age} < 60 \\
                0.5 & \text{if } 60 \leq z_\text{age} \\
 	\end{array} \right.
\end{align}
However, the assumption that the treatment effect is a step function may be too
restrictive, and $\beta(\mathbf{z})$ in reality may be a smooth interaction
function. In other words, personalised medicine is required
instead of stratified medicine.  Because methods for subgroup analyses again
generalise the treatment effect for a group of patients, it can only be considered as
a step in the direction toward personalised medicine.  We provide a method that can
estimate smooth treatment effect functions using model-based random forests and
weighted models. More importantly, this method provides an
estimate for the treatment effect of a future patient, thereby allowing a decision to be made whether treatment of this patient is appropriate.

\section{Methods} 

\cite{seibold_model-based_2015} introduced a means of
conducting subgroup analysis using model-based recursive partitioning. One first
defines a model $\mathcal{M}((Y, \mathbf{X}), \boldsymbol{\vartheta})$ with
primary endpoint $Y$, covariates $\mathbf{X}$ including the treatment indicator 
\begin{align}
  X_{A} &= \left\{ 
  \begin{array}{l l}
    1 & \quad \text{if patient received the (new) treatment}\\
    0 & \quad \text{if patient received no treatment (or standard of
	care),}
  \end{array} \right.
\end{align}
and parameter vector $\boldsymbol{\vartheta}$. In the following we will
consider likelihood models (e.g.\ generalised linear models or parametric survival models)
where the model parameters $\boldsymbol{\vartheta}$ can be estimated by
maximising the log-likelihood $l((Y, \mathbf{X}), \boldsymbol{\vartheta})$ of
those models (e.g.\ Gaussian log-likelihood or Weibull log-likelihood) or
equivalently by solving the score equation
\begin{align}
        \sum\limits_{i=1}^N 
        s((y, \mathbf{x})_i, {\boldsymbol{\vartheta}}) 
        = \mathbf{0}
\end{align}
with 
\begin{align}
	s((y, \mathbf{x})_i, {\boldsymbol{\vartheta}}) = 
	\frac{\partial l((y, \mathbf{x})_i, {\boldsymbol{\vartheta}})}{
		\partial {\boldsymbol{\vartheta}}}.
\end{align}
In most applications the model contains only an intercept $\alpha$ and a
treatment effect $\beta$, i.e.\ $\mathbf{X}=(1, X_A)$ and
$\boldsymbol{\vartheta} = (\alpha, \beta)^\top$ but more parameters are
possible, such as coefficients of additional regressors or scale and shape
parameters for the response distribution. Technically, there can also be more
than two treatment groups. For simplicity, we will focus on the simple case
with intercept and treatment effect and two treatment groups.  The method
obtains subgroups $\{\mathcal{B}_{b=1,\dots,B}\}$ that differ with regard to
the treatment effect $\beta$ and potentially the intercept $\alpha$. The
subgroups are defined by patient characteristics $\mathbf{Z} =(Z_1, \dots, Z_J)
\in \mathcal{Z}$. Hence the intercept and treatment parameters can be written
as a function of the patient characteristics $\mathbf{z}$. 

Conceptually, the partitioned model parameters $\alpha(\mathbf{z})$ and
$\beta(\mathbf{z})$ might depend on $\mathbf{z}$ in a more complex way than a
simple tree structure. Therefore, the model parameters are not step functions,
but rather smooth interaction functions, so that an individual treatment effect
(as in personalised medicine) can be computed for each patient instead of only for
each subgroup of patients (as in stratified medicine). The function
$\beta(\mathbf{z})$ can then be understood as an estimate of the individual
treatment effect of a patient with patient characteristics $\mathbf{z}$.   

The most intuitive step from a tree structure to a more complex structure is to
use a random forest instead of a single tree. Hence we propose a model-based
random forest for personalised medicine, which can be used to predict the
treatment effect of future patients using personalised models.

\subsection{Random forest} 
Random forests \citep{breiman_random_2001} compute an ensemble of $T$ trees.
The proposed algorithm draws subsamples $\mathcal{L}_t, ~t=1,\dots,T$ of the
given $N$ observations and fits a model-based tree to each subsample using a
randomly sampled set of candidate split variables $\mathbf{z}$. The data
$\mathcal{L}_t^c$ that were not in the learning sample for tree $t$ are called
out-of-bag data. 

\subsubsection*{Split procedure}
The special feature of our method is the split procedure, which is based on the 
empirical estimating function
\begin{align}\label{scoremat}
        \boldsymbol{s} &= 
        \begin{pmatrix} 
                s_{\hat\alpha}((y, \mathbf{x})_1, \hat{\boldsymbol{\vartheta}}) 
		& s_{\hat\beta}((y, \mathbf{x})_1, \hat{\boldsymbol{\vartheta}}) \\
                s_{\hat\alpha}((y, \mathbf{x})_2, \hat{\boldsymbol{\vartheta}}) 
		& s_{\hat\beta}((y, \mathbf{x})_2, \hat{\boldsymbol{\vartheta}}) \\
                \vdots & \vdots  \\
                s_{\hat\alpha}((y, \mathbf{x})_N, \hat{\boldsymbol{\vartheta}}) 
		& s_{\hat\beta}((y, \mathbf{x})_N, \hat{\boldsymbol{\vartheta}}) \\
        \end{pmatrix}
\end{align}
which contains the score contributions $s_{\hat\alpha}((y, \mathbf{x})_i,
\hat{\boldsymbol{\vartheta}})$ and $s_{\hat\beta}((y, \mathbf{x})_i,
\hat{\boldsymbol{\vartheta}})$. The score contributions are the partial
derivatives of the log-likelihood with respect to $\alpha$ or $\beta$
respectively evaluated at the $N$ observed data points and the estimated
parameters $\hat{\boldsymbol{\vartheta}} = (\hat{\alpha},
\hat{\beta})^\top$\citep{zeileis_model-based_2008}.  The matrix of score
contributions contains information on the deviation from the model fit for all
parameters and observations of a given model $\mathcal{M}((Y, \mathbf{X}),
\boldsymbol{\vartheta})$. The contributions can thus be seen as residuals. 

To obtain a split in model-based recursive partitioning for this setup the
following steps have to be performed:
\begin{itemize}
	\item Compute the prespecified (parametric) model
		 $\mathcal{M}((Y, \mathbf{X}), \boldsymbol{\vartheta})$.
	\item Compute the associated score matrix $\boldsymbol{s}$.
	\item Perform tests of independence between the residuals (partial score
	        vectors) and the partitioning variables:
        \begin{align*} 
                        H_0^{\alpha,j}&: \quad
                        {s}_{\hat\alpha}( (Y, \mathbf{X}), 
                        \hat{\boldsymbol{\vartheta}})
                        \quad\bot\quad Z_j \nonumber\\
                        H_0^{\beta,j}&: \quad
                        {s}_{\hat\beta}( (Y, \mathbf{X}), 
                        \hat{\boldsymbol{\vartheta}})
                        \quad\bot\quad Z_j  \qquad j = 1,\dots, J
         \end{align*}
 
		The smallest $p$-value corresponds
		to the greatest deviation from the model assumption, that
		intercept and treatment parameter are the same for all patients
		in the given node/subgroup.
	\item If any $p$-value is lower than the significance level, select the 
		partitioning variable that has the highest association (lowest
		$p$-value) to any of the relevant residuals for the split.
	\item Search for the optimal split point in the selected partitioning
		variable using a suitable criterion, such that the models in
		the resulting daughter-nodes have as little association between
		the partitioning variable and the residuals as possible.
\end{itemize}
This split procedure is repeated until a stop criterion is met. This can be,
for example, when no $p$-values are lower than the significance level or if
subgroups become too small. In the end a tree is obtained with disjoint
subgroups
\begin{align}
	 {\bigcup\limits_b^\bullet} \mathcal{B}_b = \mathcal{Z}
\end{align}
or a random forest with $T$ trees and disjoint subgroups for each tree
\begin{align}
        \bigcup\limits_b^\bullet \mathcal{B}_{tb} = \mathcal{Z} \quad 
	\forall t=1,\dots,T.
\end{align}

The independence tests can be performed using permutation tests
\citep{hothorn_lego_2006, hothorn_unbiased_2006} or, for reasonably large
samples, using M-fluctuation tests \citep{zeileis_generalized_2007,
zeileis_model-based_2008}. Note that multiplicity adjustment, e.g.\ Bonferroni
correction, is recommended. More details on the algorithm and the test
procedures used are documented in the appendix.

\subsection{Personalised models} 
In personalised medicine, the goal is to learn how much a person will profit
from a given treatment and what would happen if the standard of care or no
treatment is given.  For any patient, it is possible to compute a personalised
model based on the similarity of this observation to the observations in the
training data. The similarity $w_{i}(\mathbf{z})$ of a patient $i$ to any
patient from the training set is given as the number of times the patients are
classified in the same subgroup by the single trees in the random forest  
\begin{align} 
        w_{i}(\mathbf{z})&= 
	\sum\limits_{t=1}^T \sum\limits_{b=1}^{B_t} (\mathbf{z}_i \in
        \mathcal{B}_{tb}) \wedge (\mathbf{z} \in \mathcal{B}_{tb}),
\end{align}
with $T$ being the number of trees used for the computation of the forest and
$B_t$ being the number of subgroups from tree $t$
\citep{Hothorn_Lausen_Benner_2004, Meinshausen_2006}. If patient $i$ is part of
the training set, the weights can be computed out-of-bag, i.e. the only trees
($t=1,\dots,T$) considered are those where patient $i$ is not in the subset
$\mathcal{L}_t$ for the computation.

To obtain the personalised model for patient $i$ the base model is recomputed
with the weighted training data or equivalently by using the weighted score
function in the score equation
\begin{align}
	\sum\limits_{k=1}^N w_i(\mathbf{z}_k) \cdot s((y, \mathbf{x})_i,
	{\boldsymbol\vartheta}(\mathbf{z}_i)) = 0.
\end{align}
In other words every patient $k$ from the training set is included
$w_{i}(\mathbf{z}_k)$ times in the ``new data set'' to compute the personalised
model for patient $i$. In the following the parameters estimated from this
model will be denoted by $\hat{\boldsymbol\vartheta}(\mathbf{z}_i) =
(\hat{\alpha}(\mathbf{z}_i), \hat{\beta}(\mathbf{z}_i), \dots)$.

Using the personalised models it is possible to obtain a log-likelihood. From
the personalised model for patient $i$, the log-likelihood contribution
$l((y,\mathbf{x})_i , \hat{\boldsymbol{\vartheta}}(\mathbf{z}_i))$ for this
observation is computed. The log-likelihood then is 
\begin{align} 
    \sum\limits_{i=1}^N l\left((y,\mathbf{x})_i , 
    \hat{\boldsymbol{\vartheta}}(\mathbf{z}_i)\right),
\end{align}
which we refer to as forest log-likelihood.

\subsection{Improvement through personalised models}
To check whether the personalised models actually lead to an improvement
of the base model we test the hypothesis
\begin{align}
	H_0:& \nonumber \\ 
	 &\underbrace{\alpha(\mathbf{Z}) \equiv \alpha}_{H_0^\alpha} \label{eq:H0a} \\
	 & \quad \bigcap \nonumber \\
	 &\underbrace{\beta(\mathbf{Z}) \equiv \beta}_{H_0^\beta}. \label{eq:H0b}
\end{align}
This strict null hypothesis is to be rejected, if any of the patient
characteristics contain information on the outcome or the treatment effect. To
conduct the test, one can proceed as follows:
\begin{itemize}
	\item Compute the forest log-likelihood and the log-likelihood of the
		base model and calculate their difference.
	\item Draw parametric bootstrap samples from the base model.
	\item Compute the forest log-likelihood and the log-likelihood of the
		base model in the bootstrap samples and again compute the 
		differences.
	\item The $p$-value is then the proportion of bootstrap samples in which 
		the difference in log-likelihoods exceeds the observed 
		difference in the original data. Note, that this $p$-value will
		be very low or even zero when the patient 
		characteristics contain information on the outcome or the 
		treatment effect.
\end{itemize}
In practice, one may be interested in just $H_0^\beta$, but testing the
sub-hypotheses $H_0^\alpha$ and $H_0^\beta$ separately is not straight forward.
An approximation would be to compute the personalised models using a forest
that splits only based on the partial score function with respect to $\alpha$
or $\beta$ respectively. Patient characteristics, however, are often not
exclusively predictive or prognostic but may be both. Also, if a patient
characteristic is purely prognostic, this still may result in a pattern in both
partial score functions. For more details, see \cite{seibold_model-based_2015}.

\subsection{Dependence plots} 
A partial dependence plot describes the dependence of a function (in our
case the treatment effect $\hat{\beta}(\mathbf{z})$) and a variable (in our
case a partitioning variable) \citep{hastie_elements_2009}.  The partial
dependence plot resulting from a model-based tree would show a step function.
The partial dependence from a random forest can be smoother for continuous
partitioning variables. It can be obtained by plotting $\hat{\beta}({z}_j)$
against $z_{j}$ for each partitioning variable $j = 1, \dots, J$.

\subsection{Variable importance}\label{sec.vi} 
The variable importance for the random forest is computed based on the tree
log-likelihoods. For a given forest computed with $T$ trees the log-likelihood
is computed as follows:
\begin{itemize}
	\item Select the out-of-bag data $\mathcal{L}_t^c$ and determine the 
	terminal node/subgroup to which each observation $i$ belongs to.
	\item Compute the log-likelihood contribution of each observation 
	$i \in \mathcal{L}_t^c$ based on the respective model in the terminal 
	node/subgroup with parameters 
	$\hat{\boldsymbol\vartheta}(\mathbf{z}_i)$.
	\item Compute the out-of-bag log-likelihood as the sum of the
	contributions  
	\begin{align}
		l_t &=
		\sum\limits_{i \in \mathcal{L}_t^c} l((y, \mathbf{x})_i, 
		\hat{\boldsymbol\vartheta}(\mathbf{z}_i)).
	\end{align}
\end{itemize}
To obtain the variable importance of a given variable $z_j, ~j = 1,\dots,J$,
the variable is permuted.  The log-likelihood is computed as above, except that
the column with information about $z_j$ in the out-of-bag data is replaced by
the permuted $z_j$. We denote the log-likelihood of tree $t$ with variable
$z_j$ permuted by $l_t^{(j)}$. The variable importance is then
\begin{align}
	\text{VI}_j = \frac{1}{T} \sum\limits_{t=1}^T 
	\left[ l_t - l_t^{(j)}	\right].
\end{align}
If the variable importance is high the variable is an important predictive
and/or prognostic factor. Note that due to the absolute differences the 
variable importances may be negative.

\section{Results}

\setkeys{Gin}{width=0.95\textwidth}

\subsection{PRO-ACT data}
The Pooled Open Access Clinical Trials (PRO-ACT,
\url{https://nctu.partners.org/ProACT}) database contains longitudinal data of
ALS patients that participated in one of 16 phase II and III trials and one
observational study. It is a project initiated by the non-profit organisation
Prize4Life (\url{http://www.prize4life.org/}) to enhance knowledge about ALS.
It contains information on a broad variety of patient characteristics, such as
vital signs, the patient's and family's history, and treatment information.
Identification criteria, such as study centres, are not included in the
database. Also collected are the survival time and the ALS functional rating
scale (ALSFRS), which is a score measuring the patients ability of living a
normal life \citep{brooks_amyotrophic_1996}.  The ALSFRS is a sum-score of ten
items, each of which ranges between zero and four, where zero represents
complete inability and four represents normal ability. The items are speech,
salivation, swallowing, hand-writing, cutting food and handling utensils,
dressing and hygiene, turning in bed and adjusting bed clothes, walking,
climbing stairs, and breathing. As outcomes in the study we used both the
survival time (denoted by survival) and the ALSFRS six months after treatment
start (denoted by $\text{ALSFRS}_6$) and identified patient characteristics
that influence the effect of Riluzole on these outcomes.  For the two outcome
variables, we obtained two different data sets.  We only included observations
that contain information on the respective outcome variable and only patient
characteristics that have fewer than 50\% missing values.  The survival time
data set contains 3306 observations and
18 patient characteristics. The ALSFRS data set contains
2534 observations and 57 patient
characteristics.

%latex.default(tab_FRS, file = "", rowlabel = "", label = "tab.bm",     caption = "ALSFRS base model (Gaussian generalised linear \n\t\tmodel with log-link and offset). Given are the parameter \n\t\testimates, their standard error and the Wald confidence \n\t\tinterval.",     caption.loc = "top", greek = TRUE)%
\begin{table}[!tbp]
\caption{ALSFRS base model (Gaussian generalised linear 
		model with log-link and offset). Given are the parameter 
		estimates, their standard error and the Wald confidence 
		interval.\label{tab.bm}} 
\begin{center}
\begin{tabular}{lrrrr}
\hline\hline
\multicolumn{1}{l}{}&\multicolumn{1}{c}{Estimate}&\multicolumn{1}{c}{Std.Error}&\multicolumn{1}{c}{2.5 \%}&\multicolumn{1}{c}{97.5 \%}\tabularnewline
\hline
$\alpha$&$-0.1595$&$0.0065$&$-0.1722$&$-0.1468$\tabularnewline
$\beta$&$ 0.0091$&$0.0077$&$-0.0060$&$ 0.0242$\tabularnewline
\hline
\end{tabular}\end{center}

\end{table}%latex.default(tab_surv, file = "", rowlabel = "", label = "tab.bmsurv",     caption = "Survival time base model (Weibull model). Given are the \n\t\tparameter estimates, their standard error and the Wald \n\t\tconfidence interval.",     caption.loc = "top")%
\begin{table}[!tbp]
\caption{Survival time base model (Weibull model). Given are the 
		parameter estimates, their standard error and the Wald 
		confidence interval.\label{tab.bmsurv}} 
\begin{center}
\begin{tabular}{lrrrr}
\hline\hline
\multicolumn{1}{l}{}&\multicolumn{1}{c}{Estimate}&\multicolumn{1}{c}{Std.Error}&\multicolumn{1}{c}{2.5 \%}&\multicolumn{1}{c}{97.5 \%}\tabularnewline
\hline
$\alpha_1$&$ 6.7070$&$0.0323$&$ 6.6437$&$ 6.7703$\tabularnewline
$\beta$&$ 0.1073$&$0.0387$&$ 0.0314$&$ 0.1832$\tabularnewline
$\log(\alpha_2)$&$-0.5833$&$0.0271$&$-0.6364$&$-0.5302$\tabularnewline
\hline
\end{tabular}\end{center}

\end{table}

Tables \ref{tab.bm} and \ref{tab.bmsurv} show the estimates including standard
errors obtained from the base model for each outcome. For the ALSFRS this base
model is given by
\begin{align}
  \mathbb{E}\left.\left(\frac{\text{ALSFRS}_6}{\text{ALSFRS}_0} \right\vert  X
	= x\right) =
  \frac{\mathbb{E}(\text{ALSFRS}_6\vert  X = x)}{\text{ALSFRS}_0} = \exp\{ \alpha
	+ \beta x_{A}\},
\end{align}
which represents a Gaussian generalised linear model with log-link and offset
$\log(\text{ALSFRS}_0)$ where $\text{ALSFRS}_0$ is the ALSFRS that was measured
at the time of treatment start.  The base model for the survival time is given
by the Weibull model
\begin{align}
  \mathbb{P}(T \le \text{survival} | X = x) = F\left(\frac{\log(\text{survival})
	 - \alpha_1 - \beta x_A}{\alpha_2}\right),
\end{align}  
where $F$ is the cumulative distribution function of the Gompertz distribution.
Note that the Weibull model has a scale parameter in addition to the intercept,
so that both $\alpha_1$ and $\alpha_2$ control the appearance of the baseline
hazard. In the notation of Equation 4, this leads to $\mathbf{\vartheta} =
(\alpha_1, \alpha_2, \beta)^\top$.

\subsection{Personalised models}
\begin{figure}
        \centering
\includegraphics[width = 0.99\textwidth]{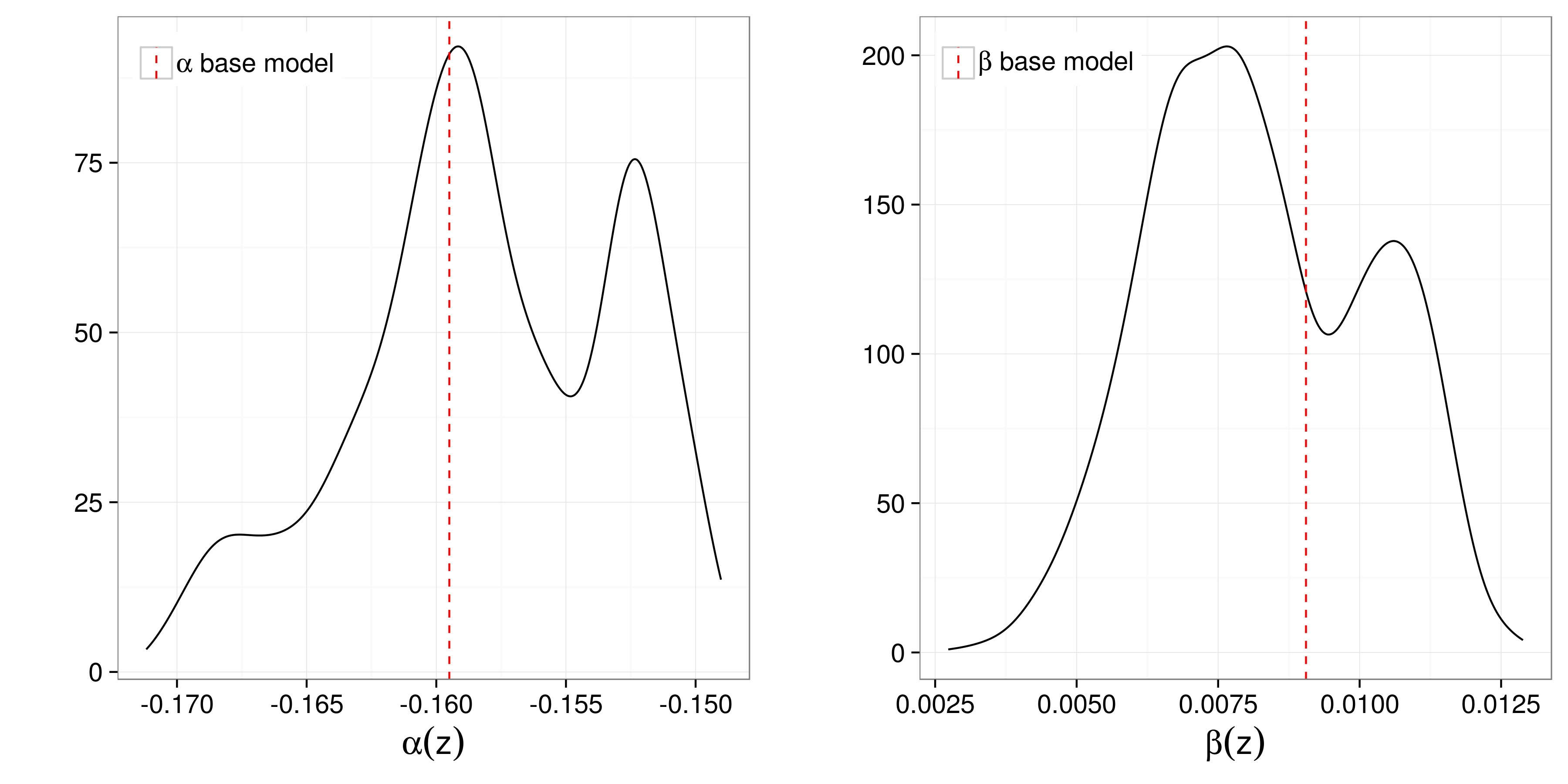}
\caption{Kernel density estimates of the personalised parameter estimates for the
         ALSFRS.}
         \label{fig.person}
\end{figure}
\begin{figure}
        \centering
	\includegraphics[width = 0.99\textwidth]{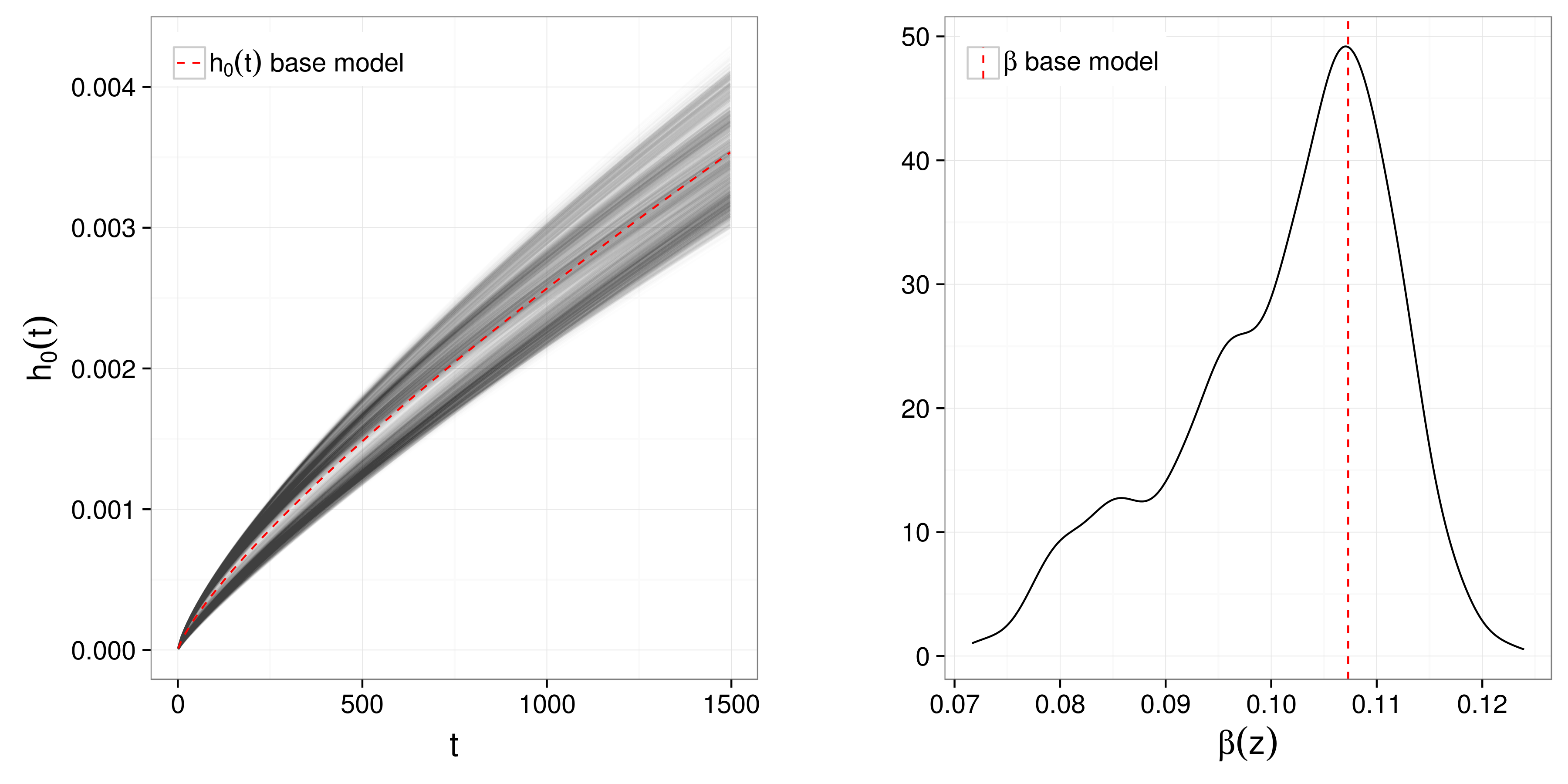}
       \caption{Distribution of the personalised parameter estimates for the
survival time. The baseline hazard functions are given in the left panel; the
kernel density estimate of the treatment effect estimate are given in the right
panel.}
        \label{fig.personsurv}
\end{figure}
We computed personalised models for all observations in the respective training
data, which were used to obtain the random forest. The distribution of
parameter estimates in the personalised models is given in
Figure~\ref{fig.person} for the ALSFRS and in Figure~\ref{fig.personsurv} for
the survival time.  Figure \ref{fig.person} shows that all patients are
predicted to have a positive Riluzole effect, i.e. for all patients taking
Riluzole, a higher ALSFRS is achieved compared to those not taking Riluzole.
However, there is a variability in the treatment effects, and the distribution
of the treatment effect is bimodal (as is the distribution of the intercept).
The treatment effect estimated from the base model is between the two modes.
The lowest treatment effect a person in this data set is predicted to have is
0.0027.  

For the survival time, the lowest predicted treatment effect is
0.0717.  However, the value of the treatment effect in the
personalised survival models cannot be interpreted in isolation; its meaning
depends on the shape of the baseline hazard, i.e. on $\alpha_1$ and $\alpha_2$.
Instead of depicting the densities of the two baseline hazard parameters, in
Figure~\ref{fig.personsurv} we show the baseline hazard curves. The baseline
hazard varies for different patients and there is a gap in the middle. The
baseline hazard estimated from the base model lies close to that gap.

From the personalised models, we obtained the ``forest log-likelihoods'' for
both outcomes. For the Gaussian GLM with log-link and offset, the log-likelihood
contribution for observation $i$ is defined as 
\begin{align} 
	l&\left((\text{ALSFRS}_6, \text{ALSFRS}_0, \mathbf{x})_i, 
	  \hat{\boldsymbol{\vartheta}}(\mathbf{z}_i) \right) \nonumber \\ 
	& \quad = 
        \left(\text{ALSFRS}_{6i} - \exp\left(\mathbf{x}_i^\top 
          \hat{\boldsymbol{\vartheta}}(\mathbf{z}_i) \right)  
	  \cdot \text{ALSFRS}_{0i} \right)^2
	%l\left((y, \mathbf{x})_i, \hat{\boldsymbol{\vartheta}}(\mathbf{z}_i)
	%\right) &=
	% (y_i - \exp(\mathbf{x}_i^\top 
	%\hat{\boldsymbol{\vartheta}}^*(\mathbf{z}_i)))^2
\end{align}
with $\mathbf{x}_i = (1, x_{Ai})^\top$ and
$\hat{\boldsymbol{\vartheta}}(\mathbf{z}_i) =
 (\hat{\alpha}(\mathbf{z}_i), \hat{\beta}(\mathbf{z}_i))^\top$.
For the Weibull model the log-likelihood contribution for observation $i$ is
\begin{align}
  l&\left((\text{survival}, \mathbf{x})_i, 
  \hat{\boldsymbol{\vartheta}}(\mathbf{z}_i)\right) \nonumber \\
	& \quad =
	\delta_i  \log(\hat{\alpha}_2(\mathbf{z}_i)) - \delta_i 
	\frac{\text{survival}_i - \mathbf{x}_i^\top 
	\hat{\boldsymbol{\vartheta}}^*(\mathbf{z}_i)}
	{\hat{\alpha}_2(\mathbf{z}_i)} + 
	\exp\left( \frac{\text{survival}_i -\mathbf{x}_i^\top 
	\hat{\boldsymbol{\vartheta}}^*(\mathbf{z}_i)}
	{\hat{\alpha}_2(\mathbf{z}_i)}\right)
\end{align}
with $\mathbf{x}_i = (1, x_{iA})^\top$,
$\hat{\boldsymbol{\vartheta}}^*(\mathbf{z}_i) = (\hat{\alpha}_1(\mathbf{z}_i),
\hat{\beta}(\mathbf{z}_i))^\top$ and $\delta_i$ as the censoring indicator.

\begin{figure}
        \centering
\includegraphics{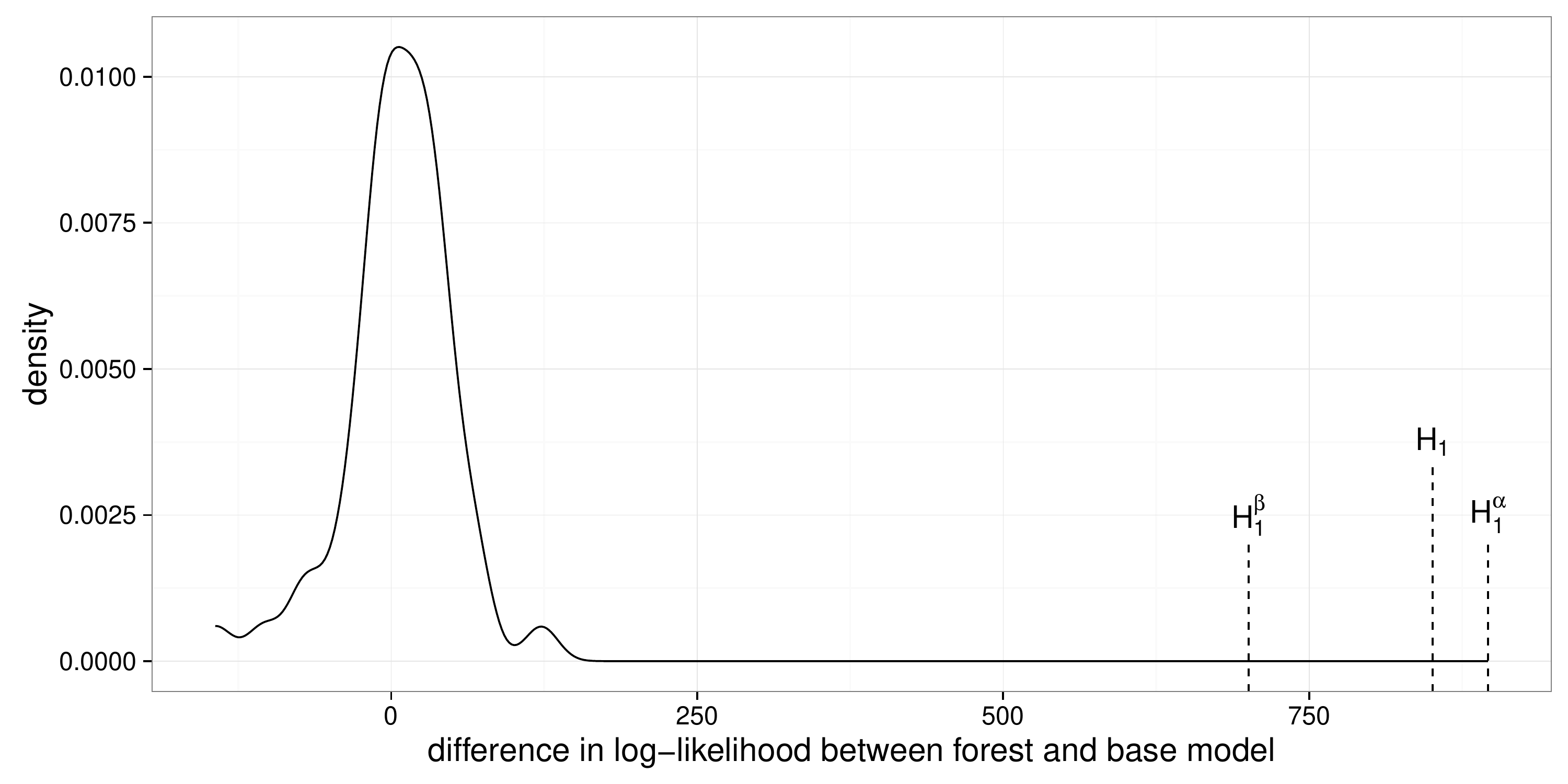}
\caption{Difference in log-likelihoods between forest and base model on the
original data (dashed lines; $H_1$, the usual forest; $H_1^\alpha$ the forest
that splits based on $\alpha$; $H_1^\beta$ the forest that splits based on
$\beta$) and on 50 samples simulated from the base model (density curve) for
the ALSFRS outcome.}
\label{fig.logliksALS}
\end{figure}

\begin{figure}
        \centering
\includegraphics{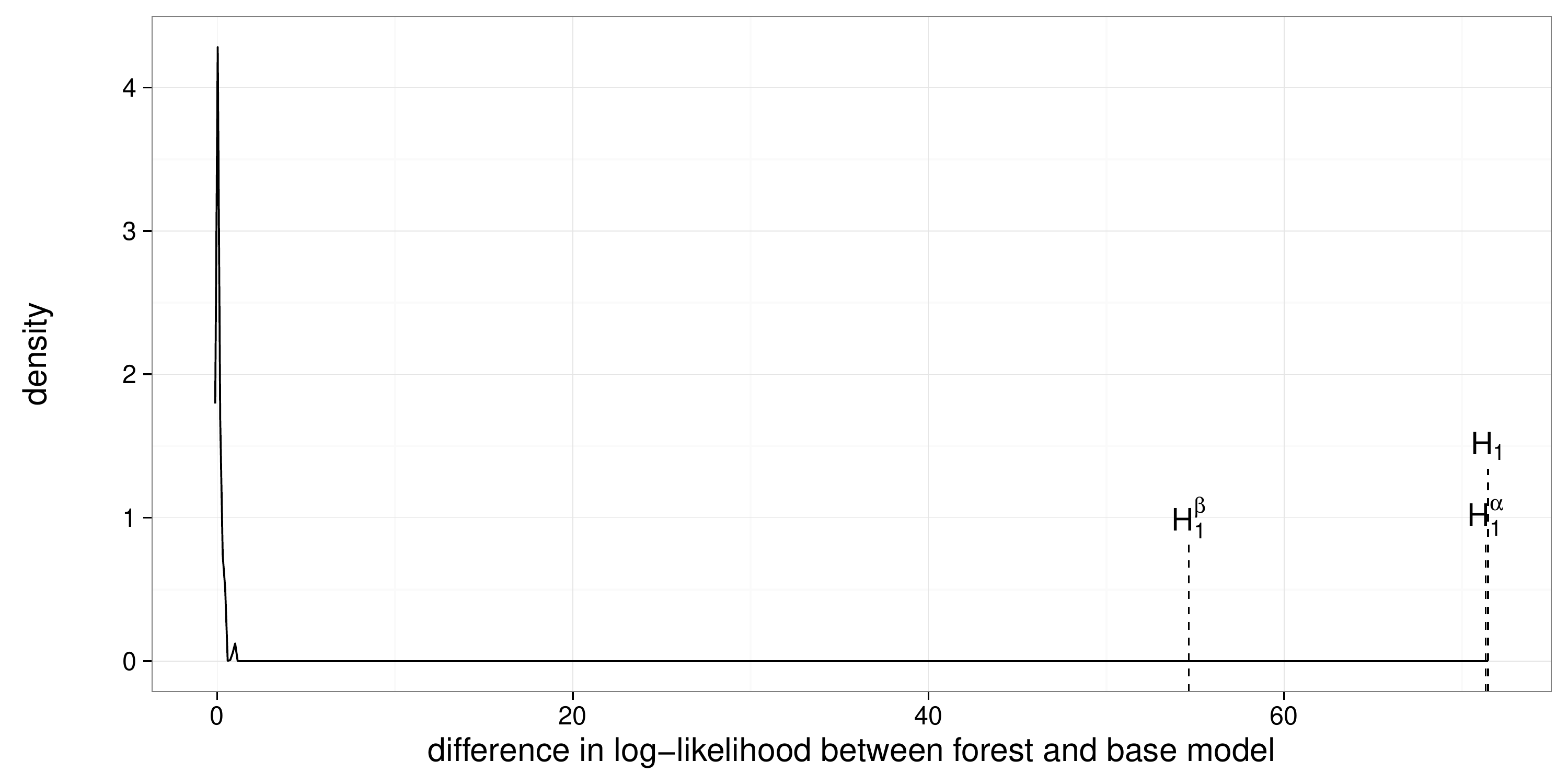}
\caption{Difference in log-likelihoods between forest and base model on the
original data (dashed lines; $H_1$, the usual forest; $H_1^\alpha$ the forest
that splits based on $\alpha$; $H_1^\beta$ the forest that splits based on
$\beta$) and on 50 samples simulated from the base model (density curve) for
the survival outcome.}
\label{fig.loglikssurv}
\end{figure}

As can be seen in Figures~\ref{fig.logliksALS} and \ref{fig.loglikssurv} the
forest log-likelihoods are higher than the log-likelihoods of the base models
for both the ALSFRS and the survival time. The figures show the difference in
log-likelihood between the forest and the corresponding base model. To show
that this difference is not due to overfitting, we drew 50 samples from the
base models, i.e.\ 50 parametric bootstrap samples for which the assumption
holds that the intercept (or baseline hazard) and treatment effect are the same
for all patients.  ALSFRS values are drawn from a normal distribution truncated
at zero to assure positivity. (The effect of truncation is virtually
negligible; only two observations had a truncation probability of more than
1\%.) The survival times are drawn from a Weibull distribution censored at the
originally observed censoring times (if exceeded).  The difference in
log-likelihoods for both ALSFRS and survival time are distributed close to
zero, with a slight shift to the right, for the parametric bootstrap samples.
The large difference in the ALS data supports the assumption that the base
models are not ideal and personalised models are meaningful (the respective
p-values are both zero). To approximately check the sub-hypotheses given in
equation \ref{eq:H0a} and \ref{eq:H0b}, we also computed log-likelihoods of the
two forests that split only with respect to one of the partial score functions
-- either intercept (or baseline hazard) or treatment effect. For the ALSFRS,
both the forest under $H_1^\alpha$ (computed with only splitting based on the
partial score function with respect to the intercept $\alpha$), as well as the
forest under $H_1^\beta$ (computed with only splitting based on
partial score function with respect to the treatment effect $\beta$), greatly
improved compared to the base model.  The difference in log-likelihood between
the forest under $H_1^\alpha$ and the base model is even greater than between
the original forest ($H_1$) and the base model. For the survival time, the
log-likelihoods of the original forest and the forest under $H_1^\alpha$ (based
on splits in the partial score function with respect to the baseline hazard)
are very close to each other.  Splitting only based on the partial score
function with respect to the treatment effect ($H_1^\beta$) already improves
the log-likelihood but not as much as splitting based on both intercept and
treatment effect ($H_1$). The good performance of the forests under
$H_1^\alpha$ indicate that (1) there are no predictive patient characteristics,
(2) all predictive patient characteristics are also prognostic, or (3) the
predictive nature of the predictive patient characteristics are so strong that
it has enough impact on the structure of the partial score function with
respect to $\alpha$.

\subsection{Dependence plots}

\begin{figure}
        \centering
        \begin{subfigure}[t]{0.49\textwidth}
\includegraphics{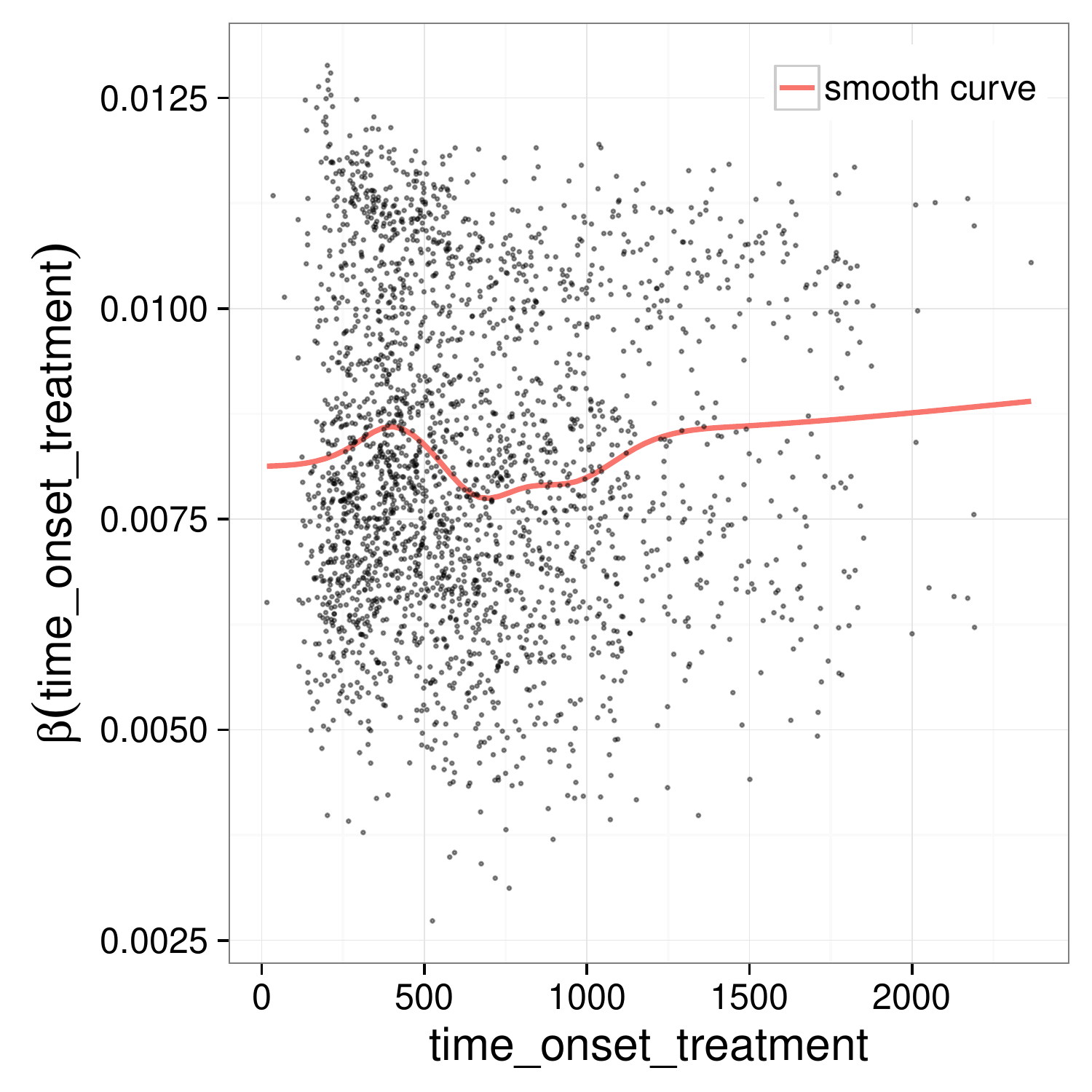}
        \caption{Dependence plot for the time in days between disease onset and treatment start.}
        \label{fid.dptime_onset_treatment}
        \end{subfigure} 
        \begin{subfigure}[t]{0.49\textwidth}
\includegraphics{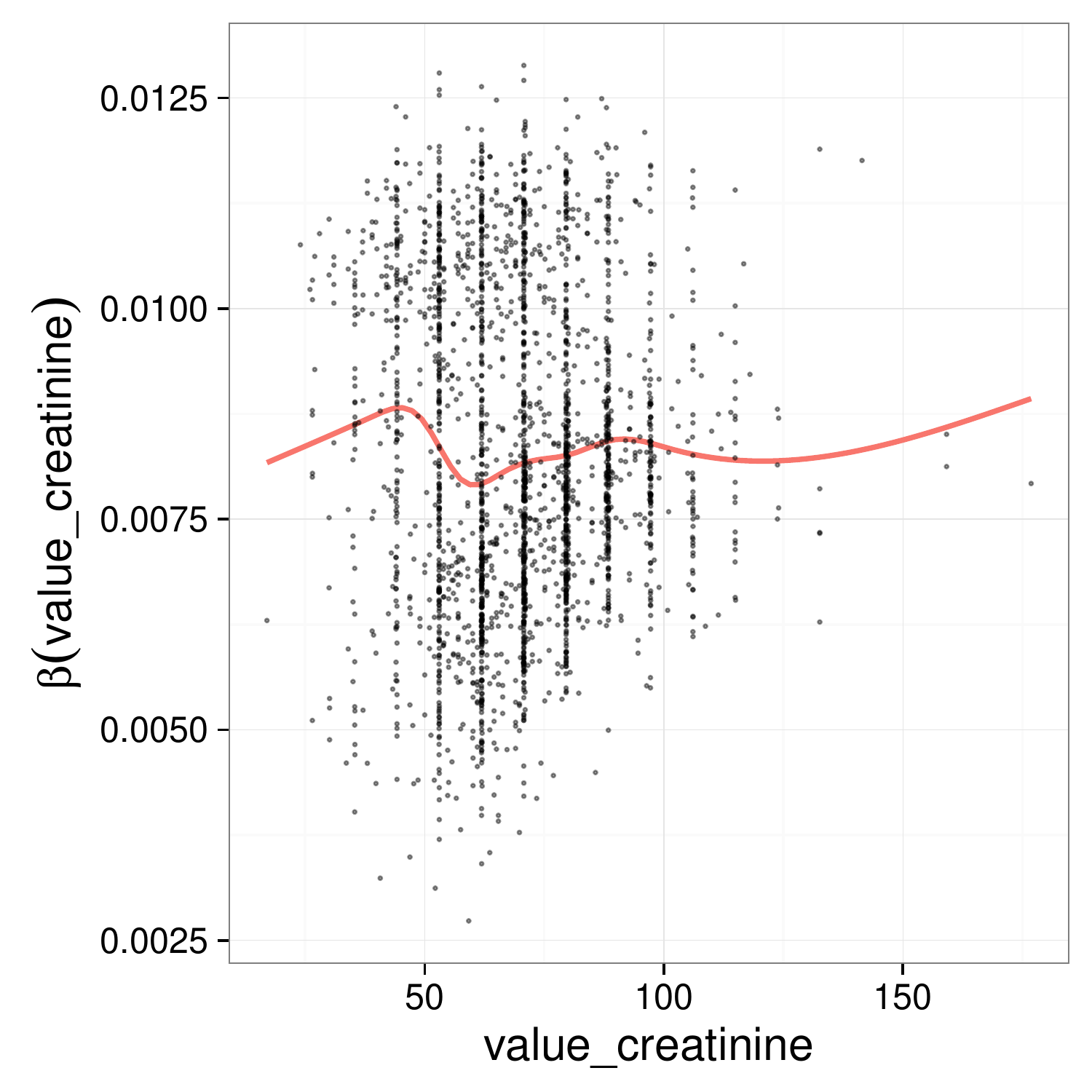}
        \caption{Dependence plot for the creatinine level in mmol/L.}
        \label{fid.dpvalue_creatinine}
        \end{subfigure} 
        \begin{subfigure}[t]{0.49\textwidth}
\includegraphics{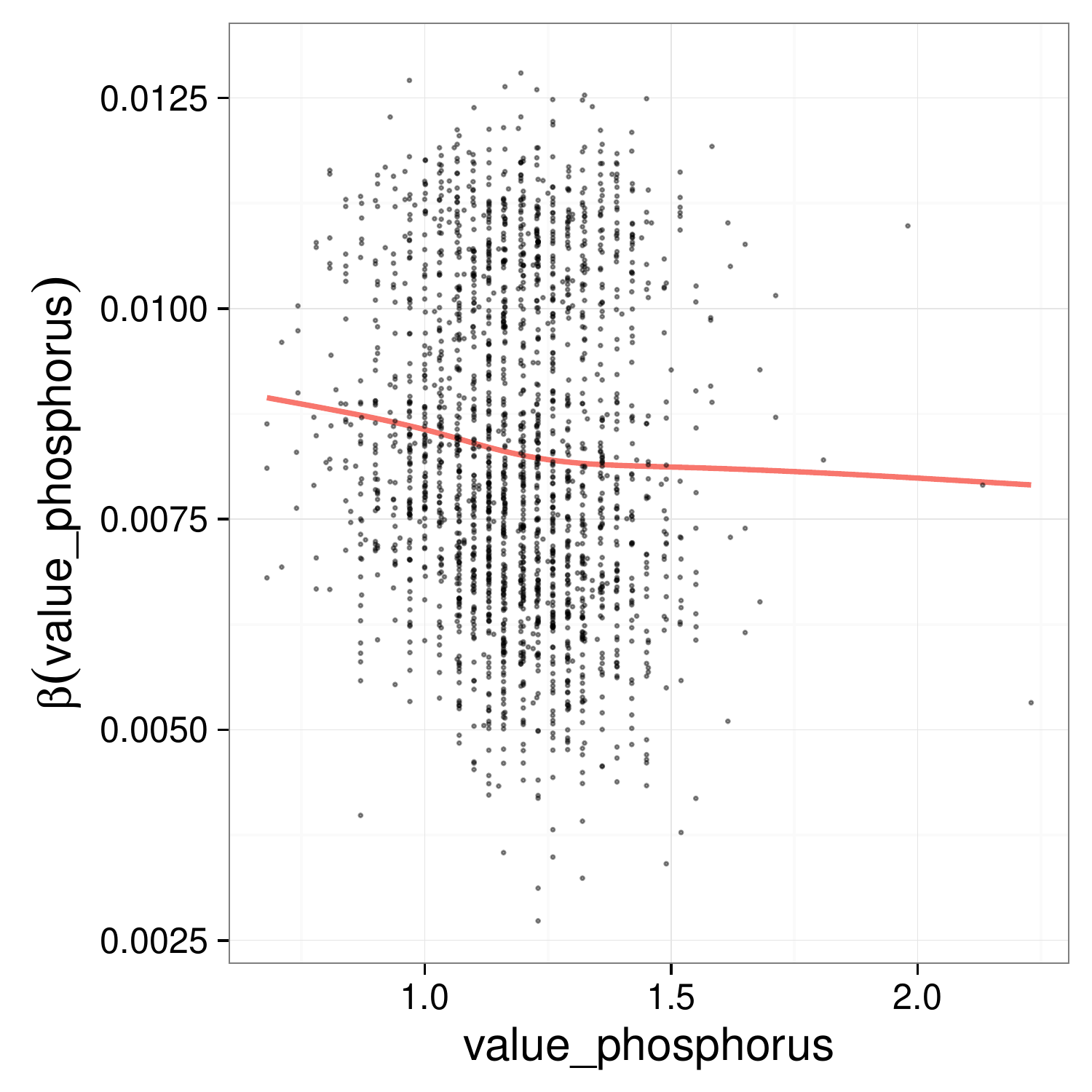}
        \caption{Dependence plot for the phosphorus level in mmol/L.}
        \label{fid.dpvalue_phosphorus}
        \end{subfigure} 
        \begin{subfigure}[t]{0.49\textwidth}
\includegraphics{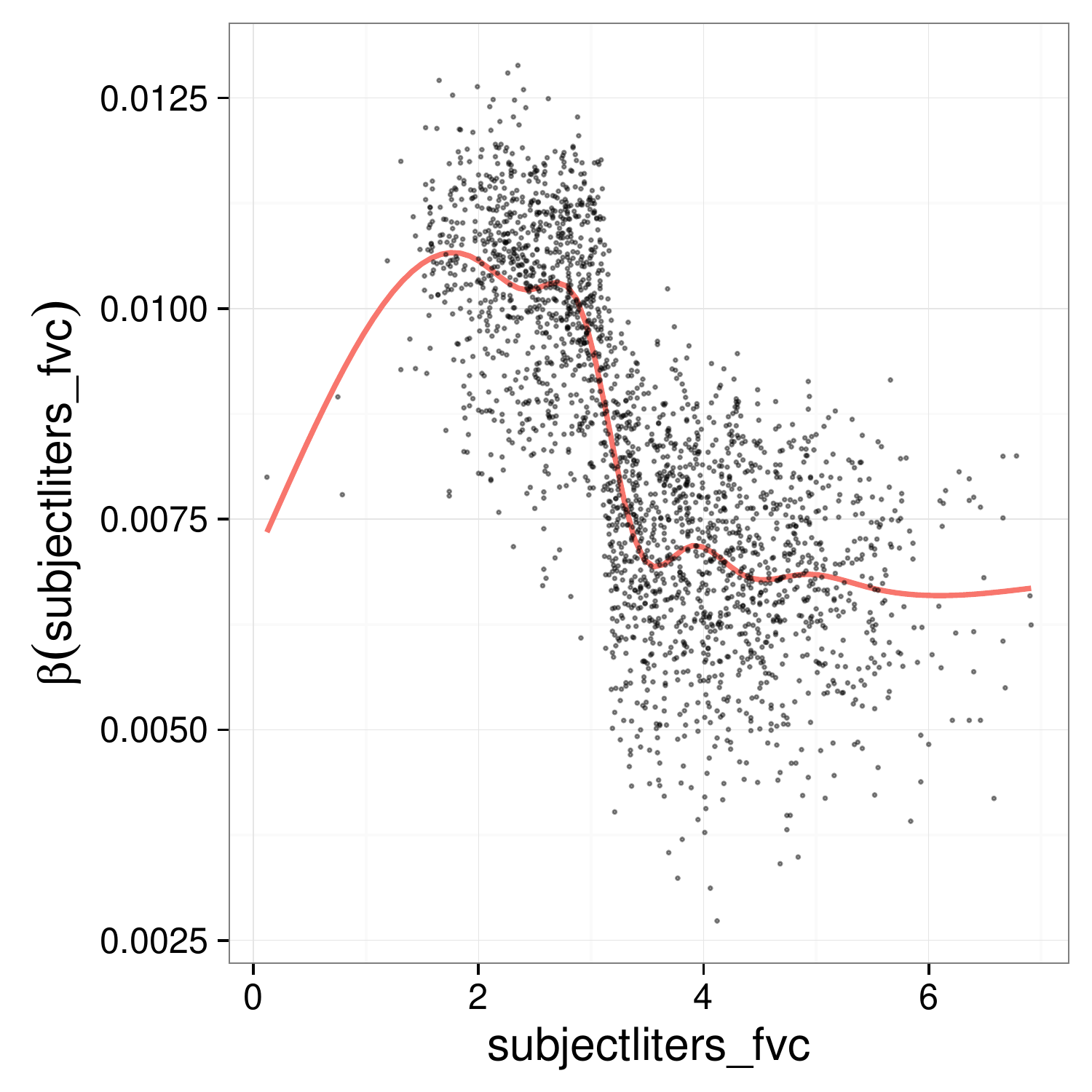}
        \caption{Dependence plot for the forced vital capacity (volume of air in litres that can forcibly be blown out after full inspiration).}
        \label{fid.dpsubjectliters_fvc}
        \end{subfigure} 
\caption{Dependence plots for the four patient characteristics with the highest
variable importance from the ALSFRS forest.}
\label{fig.dpALS}
\end{figure}

\begin{figure}
        \centering
        \begin{subfigure}[t]{0.49\textwidth}
\includegraphics{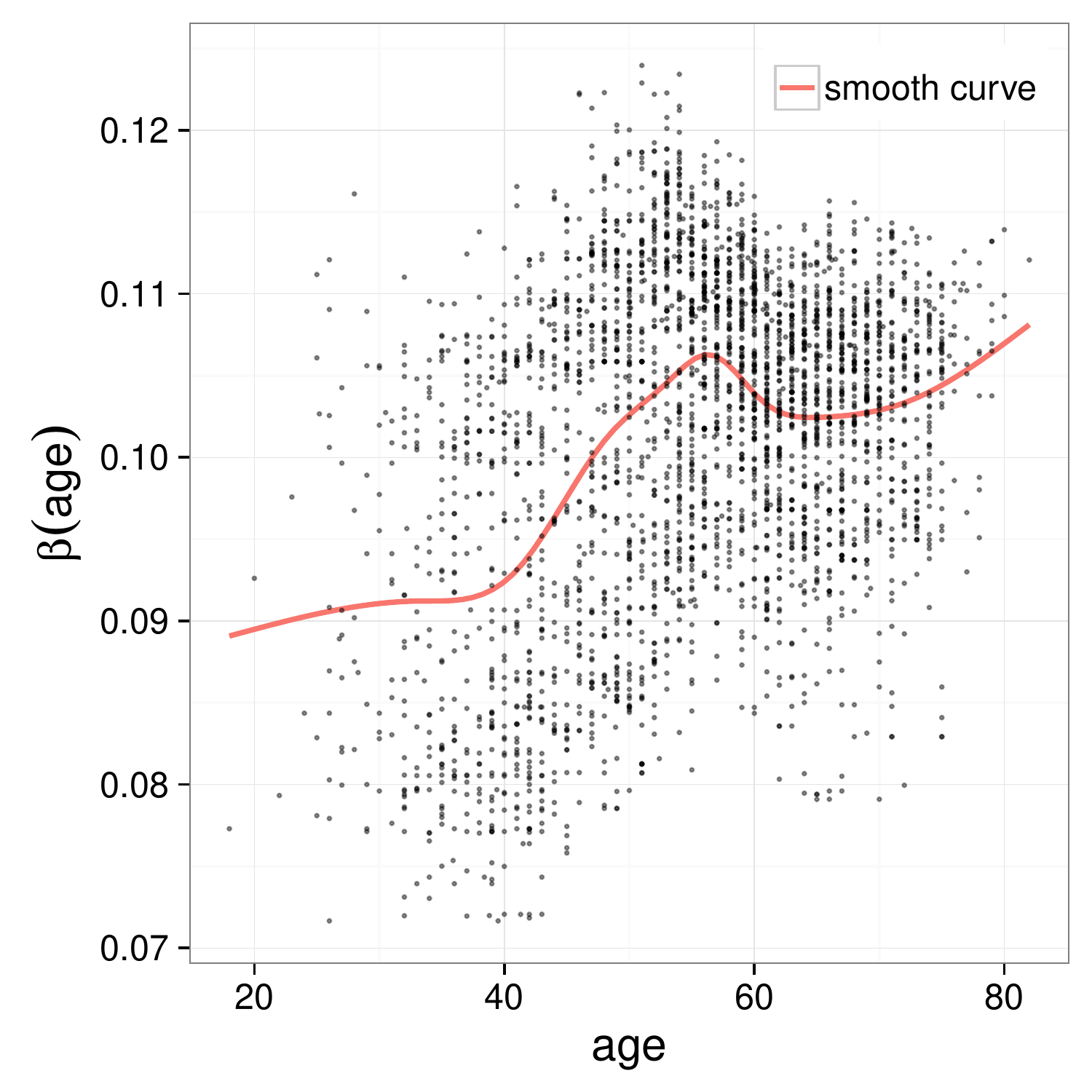}
        \caption{Dependence plot for the age.}
        \label{fid.dpsurvage}
        \end{subfigure} 
        \begin{subfigure}[t]{0.49\textwidth}
\includegraphics{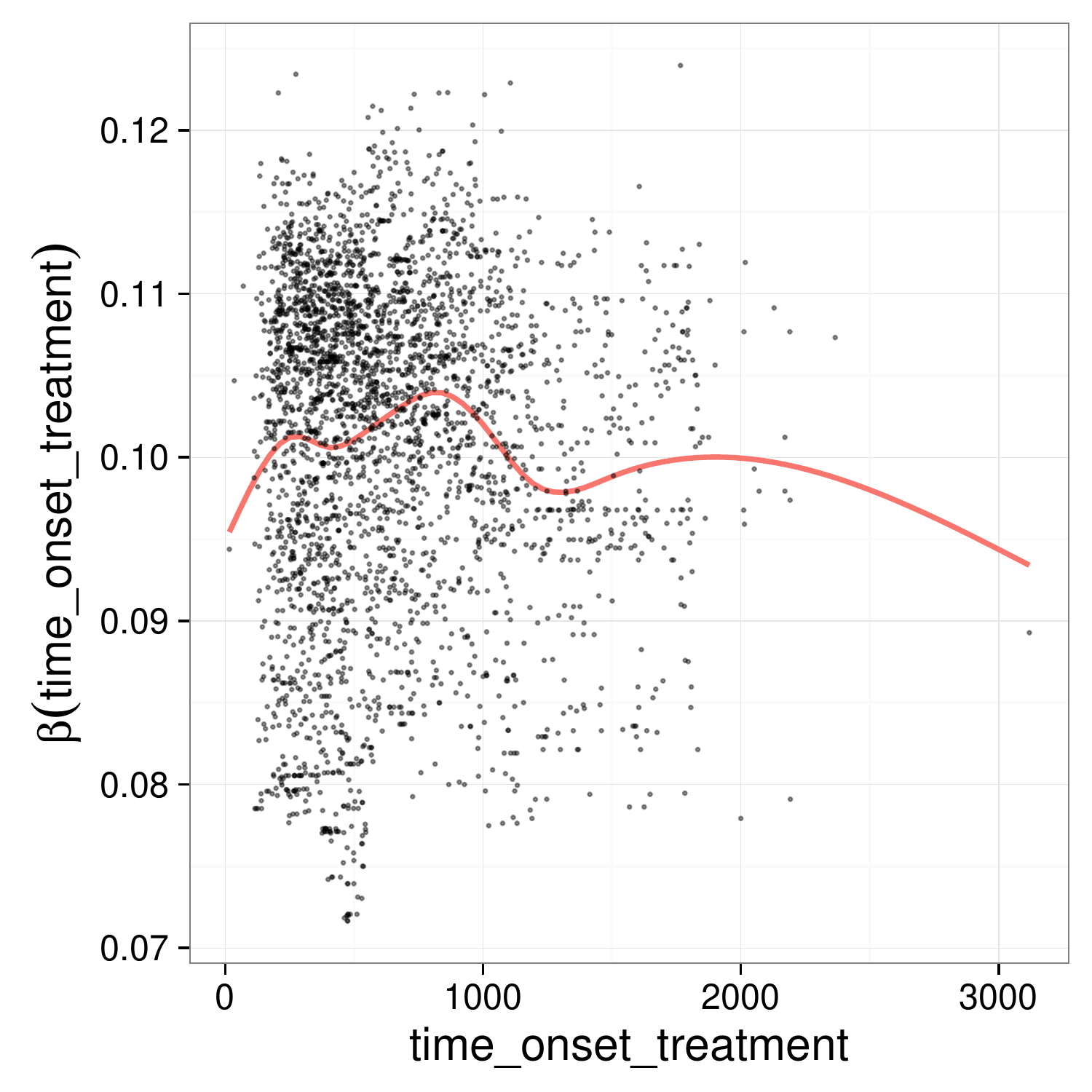}
        \caption{Dependence plot for the time in days between disease onset and treatment start.}
        \label{fid.dpsurvtime_onset_treatment}
        \end{subfigure} 
        \begin{subfigure}[t]{0.49\textwidth}
\includegraphics{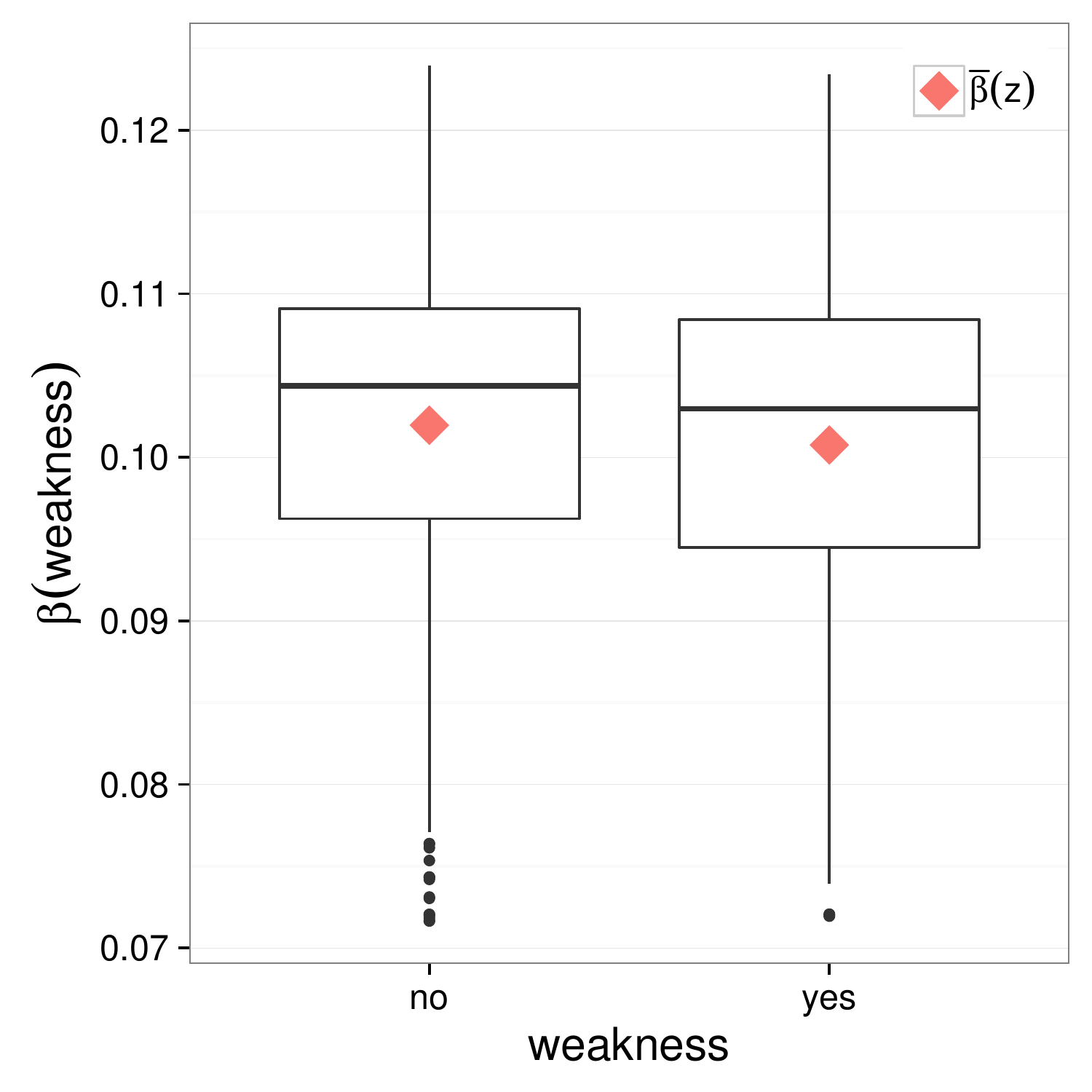}
        \caption{Dependence plot for the weakness indicator.}
        \label{fid.dpsurvweakness}
        \end{subfigure} 
        \begin{subfigure}[t]{0.49\textwidth}
\includegraphics{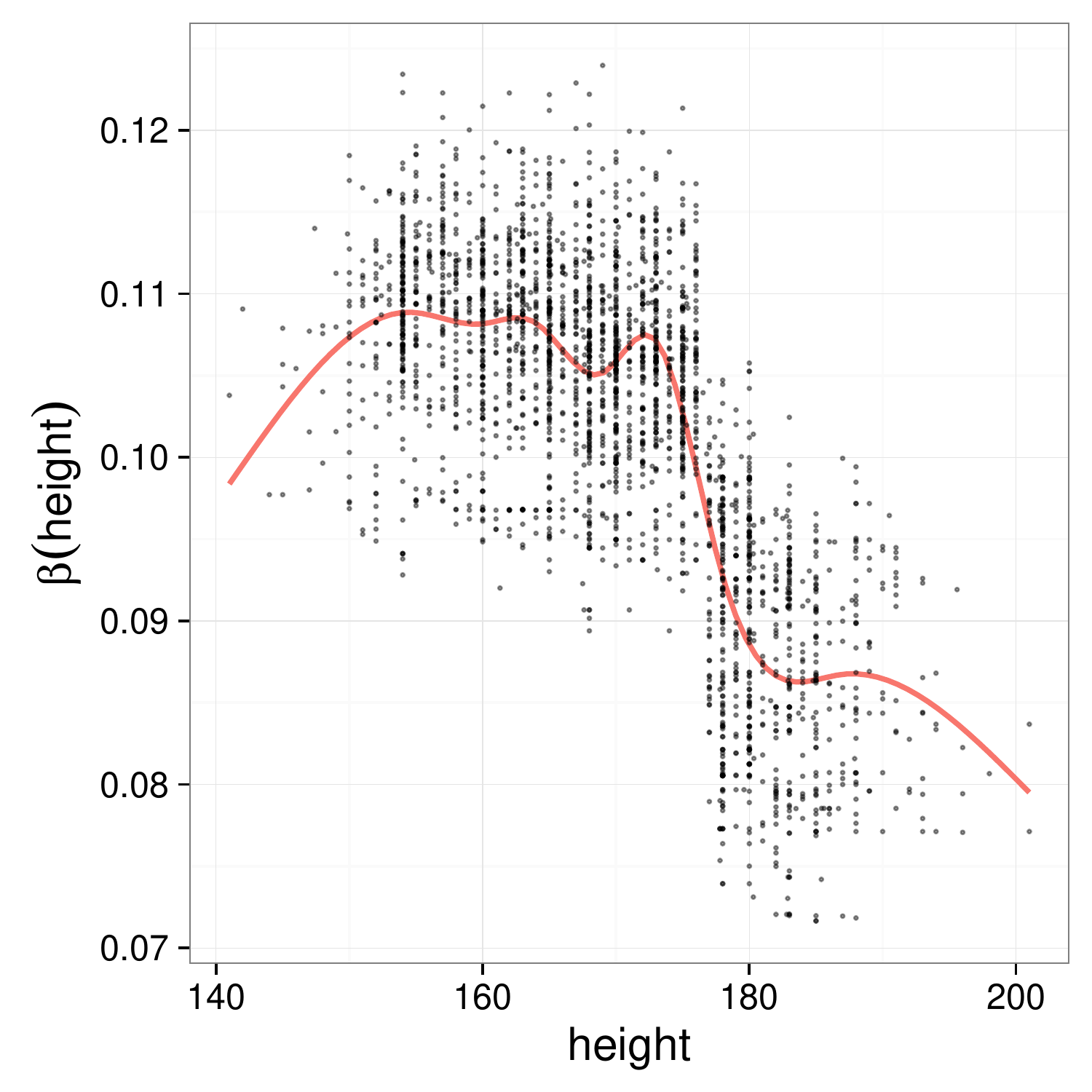}
        \caption{Dependence plot for the height.}
        \label{fid.dpsurvheight}
        \end{subfigure} 
\caption{Dependence plots for the four patient characteristics with the highest
variable importance from the survival time forest.}
\label{fig.dpsurv}
\end{figure}

The dependence plots as shown in Figures~\ref{fig.dpALS} and \ref{fig.dpsurv}
can be obtained for any partitioning variable. Here we show the dependence
plots for the four variables with the highest variable importance (see
Section~\ref{sec.resvi}). For continuous variables, such as age, we show a
scatter plot, as before. For categorical variables, such as the variable
weakness, which indicates whether a patient suffers from muscle weakness
(yes/no), boxplots giving the variation of $\beta(z)$ and a square representing
$\bar{\beta}({z})$, i.e. the mean, are a meaningful way of representing the
dependence between treatment effect and the given variable.

The most obvious pattern of the four graphs in Figure~\ref{fig.dpALS} is shown
in Subfigure~\ref{fid.dpsubjectliters_fvc}, in which the personalised treatment
effects are plotted against the forced vital capacity (FVC). Patients with a
low lung function (low FVC) are predicted to have a higher treatment effect
than those with better lung function. The graph shows a relatively clear cut at
approximately three litres. This indicates that FVC is a predictive factor. For
the time between disease onset and treatment start, the pattern is less clear.
Patients with a short as well as those with a long time between disease onset
and treatment start seem to benefit most. Also for the creatinine value, which
indicates kidney function, only weak patterns are observed. The phosphorus
balance is slightly negatively associated with the treatment effect. 

For the survival time, plotting only the treatment effect against a variable is
not meaningful since the interpretation of the treatment effect depends on the
shape of the baseline hazard. Therefore, we took a different approach in this
case and show on the $y$-axis the difference in median survival between treatment
and control intake. For example, a value of 70 means that based on the
personalised model of this patient, the median survival is prolonged by 70 days
if the patient takes Riluzole. The difference in median survival is denoted by
$\Delta_{0.5}$. Any other quantile could be used as well since from the Weibull
model, information on the entire estimated distribution in the two treatment
groups is obtained. Taking the difference in medians makes sense because it is
a measure on the scale of the outcome, just as the treatment effect in a linear
model, which is the difference in means.  The shape of $\Delta_{0.5}$ when
plotted against age shows a strong pattern that indicates that age is a
predictive factor (see Figure~\ref{fig.dpsurv}).  The treatment efficacy
increases with age until about 55 years and then flattens The difference in
median survival slightly increases with the days between disease onset and
start of treatment in the beginning, but decreases again after about 1000 days.
Patients who suffer from weakness have a greater variance in their benefit from
Riluzole.  Tall patients are predicted to benefit little on average.

\subsection{Variable importance}
\label{sec.resvi}

\begin{figure}
        \centering
        \includegraphics[width=0.8\textwidth]{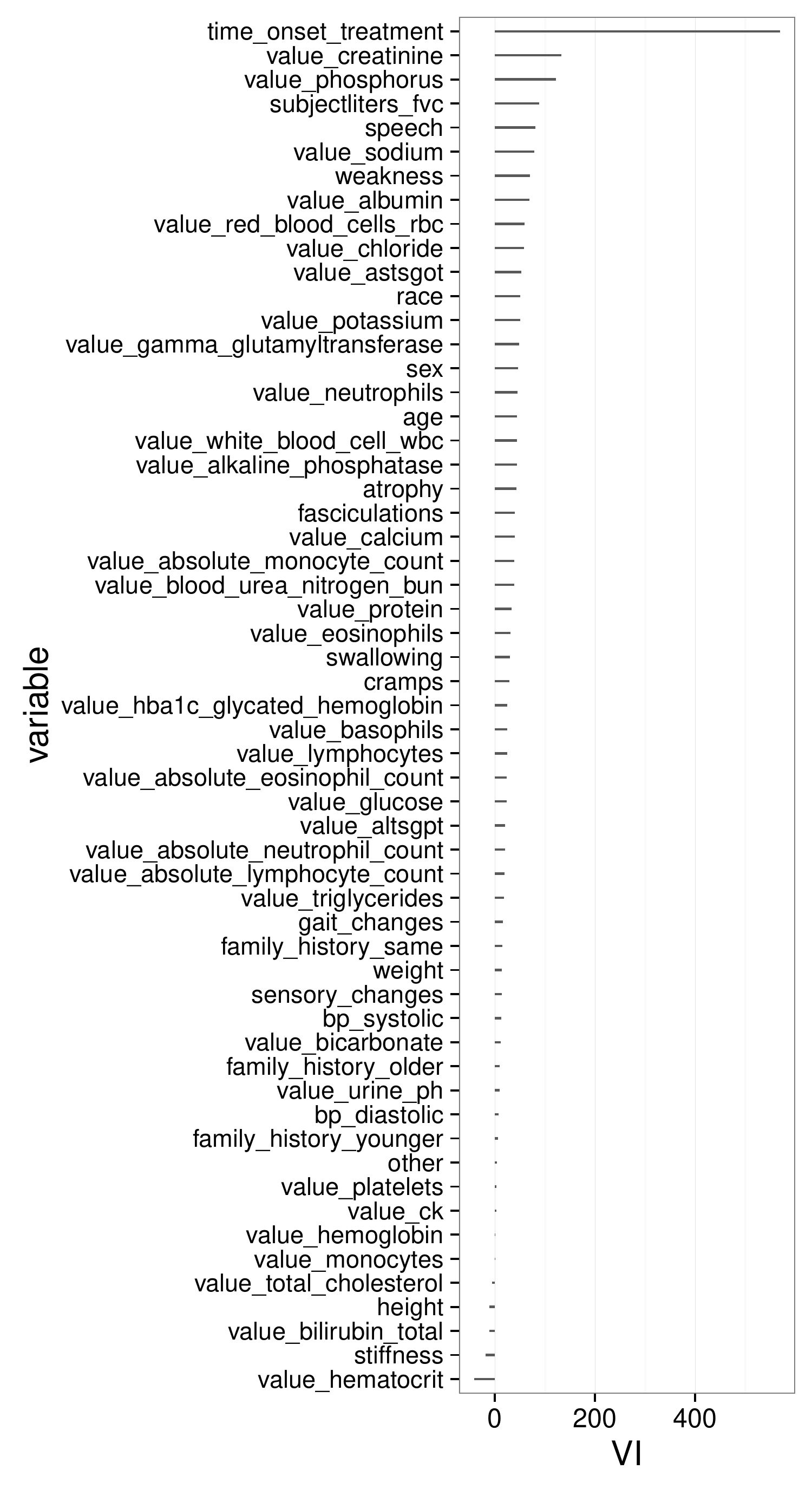}
        \caption{Variable importances of all split variables used for the 
                 ALSFRS forest.}
        \label{fig.vi}
\end{figure}

\begin{figure}
        \centering
        \includegraphics[width=0.8\textwidth]{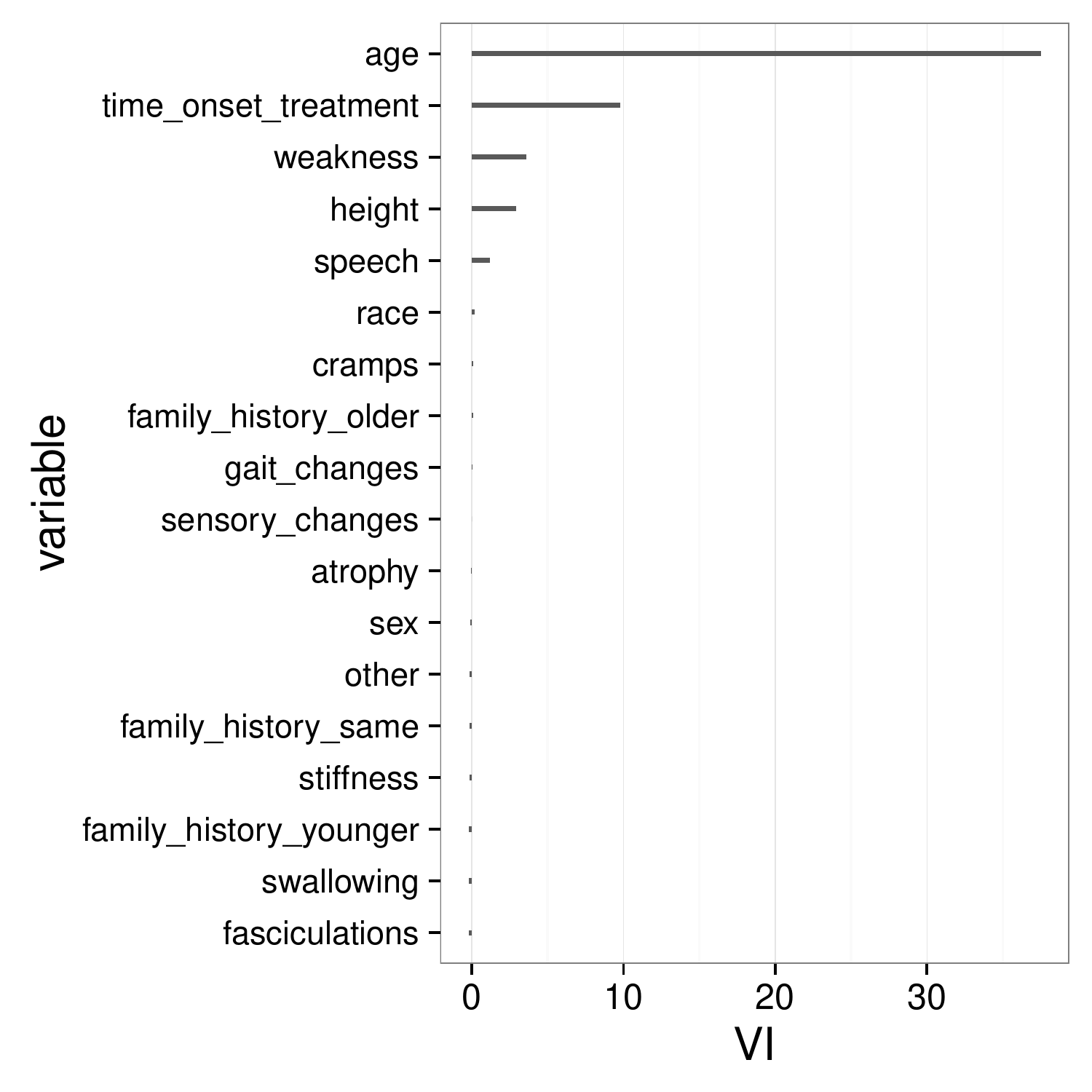}
	\caption{Variable importances of all split variables used for the
		 survival time forest.}
        \label{fig.visurv}
\end{figure}

Figures \ref{fig.vi} and \ref{fig.visurv} show the variable importance of each
split variable. Figure \ref{fig.vi} suggests that the time between disease
onset and start of treatment plays the most important role for the personalised
models.  The time between disease onset and start of treatment, the forced
vital capacity (FVC), and the phosphorus balance have been shown to be the most
important variables for stratified models \citep{seibold_model-based_2015}
which is underlined by this analysis. The time between disease onset and start
of treatment contains information on the state of disease progression for
patients in the trial. If the disease onset and the start of treatment are far
apart, the patient is likely to have a slow progression
\citep{hothorn_randomforest4life:_2014}. Also Riluzole has been shown to not be
effective when the disease is already far progressed \citep{ema_riluzole_2012}.
Thus it is not surprising that this variable is selected as an important
variable.  

For the Riluzole effect on the survival time the patient's age and again the
time between onset and treatment start play a role.  Both variables have been
identified before \citep{seibold_model-based_2015} as important factors
for survival time.

\section{Discussion}
Model-based forests can find important predictive and prognostic patient
characteristics and -- more importantly -- via the personalised models provide
the possibility to predict the treatment effect of a future patient.  Through
analysis of the PRO-ACT data and simulations (see appendix \ref{app_eval}), we
showed that personalised models can perform better than the standard global 
model if there are differences in treatment effect between patients. If there
is no difference, the performance of the methods is about the same. These
results allow a shift from standardised medicine back to personalised medicine,
but this time in a controlled way by using statistical principles.

The presented methods are based on tree-based subgroup analyses but go a step
further. Not only are subgroups identified and the treatment effect within each
group estimated, but many slightly varying trees are used to retrieve a measure
of similarity between patients. On this basis, a model is computed in which
more similar patients are weighted higher. The personalised models provide
point estimates for the treatment effect. When the individual treatment effects
are plotted against patient characteristics researchers can determine on
whether the patient characteristics are predictive factors and in what way the
patient characteristics and the treatment effect are interacting. For ALS
patients, the FVC value was predictive on the ALSFRS, and the patient's age and
height were predictive on survival. The next step would be to generate
hypotheses from these findings and plan a study to test these. Our method
offers a promising means of providing individual treatment effect predictions
and can be applied to any clinical trial data where baseline patient
characteristics are available.

All results were obtained solely using open-source implementation software (see
Section~\ref{sec.cd}), which provides easy access to the methods.

\section{Computational details}\label{sec.cd}
The code for data preprocessing of the PRO-ACT data is available in the TH.data
package \citep{hothorn_data_2014}. The source code for the full analyses is
available on \url{https://github.com/HeidiSeibold/personalised_medicine}.
Implementation of all methods discussed in this paper are based on the R
partykit package \citep[][version 1.0-2]{hothorn_partykit_2015}. Other R
packages used were sandwich \citep[][2.3-3]{zeileis_econometric_2004,
zeileis_object_2006}, survival \citep[][2.38-1]{therneau_survival_2015}, eha
\citep[][2.4-2]{brostrom_eha_2014} and ggplot2
\citep[][2.0.0]{wickham_ggplot2_2009}. All computations were conducted in the R
system for statistical computing \citep[][version 3.2.0]{R}.

\section*{Acknowledgements}
We thank the Swiss National Fund for funding this
project with grant 205321\_163456.

%% \clearpage

\bibliography{whatif} %%%,packages}

\newpage

\clearpage

\begin{appendix}

\section{Split algorithm in detail}
In the following, the split algorithm in model-based recursive partitioning is
explained. The split procedure starts with all $N$ data points. In nodes other
than the root node, the size of the data set depends on the previous splits.
For notational simplicity, we describe the split procedure in the root 
node, i.e.\ for patients $i = 1,\dots,N$. 
\begin{itemize}
 \item Compute prespecified (parametric) model
       	$\mathcal{M}((Y, \mathbf{X}), \boldsymbol{\vartheta})$. 
	Estimate $\hat{\boldsymbol{\vartheta}}$ by maximising the
	log-likelihood 
	$$\hat{\boldsymbol{\vartheta}} = 
	\operatornamewithlimits{argmax}\limits_{\vartheta} 
	l((Y, \mathbf{X}), \boldsymbol{\vartheta}) $$
	or equivalently by solving
	$$\sum\limits_{i=1}^N s((y, \mathbf{x})_i, 
	{\boldsymbol{\vartheta}}) = 0 $$
	for $\boldsymbol{\vartheta}$.
 
 \item Compute associated empirical estimating function (residuals)
        $$
	\boldsymbol{s} = 
        \begin{pmatrix} 
                s_{\hat\alpha}((y, \mathbf{x})_1, \hat{\boldsymbol{\vartheta}}) 
                & s_{\hat\beta}((y, \mathbf{x})_1, \hat{\boldsymbol{\vartheta}}) \\
                s_{\hat\alpha}((y, \mathbf{x})_2, \hat{\boldsymbol{\vartheta}}) 
                & s_{\hat\beta}((y, \mathbf{x})_2, \hat{\boldsymbol{\vartheta}}) \\
                \vdots & \vdots  \\
                s_{\hat\alpha}((y, \mathbf{x})_N, \hat{\boldsymbol{\vartheta}}) 
                & s_{\hat\beta}((y, \mathbf{x})_N, \hat{\boldsymbol{\vartheta}}) \\
        \end{pmatrix}
	$$

  \item Perform tests of independence between residuals (partial score
        vectors) $\boldsymbol{s}_{\hat\alpha}$ as well as
        $\boldsymbol{s}_{\hat\beta}$ and partitioning variables $Z_j$.
        \begin{align*} 
                        H_0^{\alpha,j}&: \quad
                        \boldsymbol{s}_{\hat\alpha}( (Y, \mathbf{X}), 
                        \hat{\boldsymbol{\vartheta}})
                        \quad\bot\quad Z_j \nonumber\\
                        H_0^{\beta,j}&: \quad
                        \boldsymbol{s}_{\hat\beta}( (Y, \mathbf{X}), 
                        \hat{\boldsymbol{\vartheta}})
                        \quad\bot\quad Z_j  \qquad j = 1,\dots, J
         \end{align*}
	For the tests, we use permutation testing with the linear statistic
	$$
	T_j = \sum\limits_{i \in \mathcal{B}_b} g_j(Z_{ji}) \cdot \mathbf{s}_i 
	$$
	The transformation function $g$ depends on the scale of the variable
	$Z_j$. If $Z_j$ is numeric then $g_j(z_{ji}) = z_{ji}$. If $Z_j$ is 
	categorical with $K$ categories then $g_j(z_{ji}) = e_K(z_{ji}) =
	(I(z_{ji} = 1), \dots, I(z_{ji} = K))$, i.e., 
	$g_j$ is the unit vector of length $K$, where the element, that
	corresponds to the value of $z_{ji}$, is one. If there are missing values in 
	$Z_j$ the observations are excluded from the sum so that we actually
	sum over all observations $i \in \mathcal{B}_b$, except for the observations
	in $\mathcal{B}_b$, where $Z_j$ is missing.
	The standardised test statistic is the Pearson correlation coefficient
	$$
        c({t}_j, \mu_j, \Sigma_j) = 
        \left| \frac{({t}_j - \mu_j)}{\sqrt{(\Sigma_j)}} \right|
	$$
	if $Z_j$ is numeric and otherwise 
	$$
	c(\mathbf{t}_j, \mu_j, \Sigma_j) = 
	\max\limits_{k = 1,\dots,K}
	\left| \frac{(\mathbf{t}_j - \mu_j)_k}{\sqrt{(\Sigma_j)_{kk}}} \right|
	.$$
	The conditional expectation $\mu_j$ and covariance $\Sigma_j$ can be 
	derived as in \cite{strasser_asymptotic_1999}. 
	The smallest $p$-value corresponds to the largest discrepancy from the
	model assumption, that intercept and treatment parameter are the same 
	for all patients in the given node/subgroup. 
  \item If any Bonferroni adjusted $p$-value is lower than the significance
	level, select the partitioning variable $Z_{j*}$ that has the highest 
	association (lowest $p$-value) to any of the residuals relevant for
	the split.
  \item Select as split point the point that results in the largest 
	discrepancy between score functions in the two resulting subgroups.
	The discrepancy can be measured by the linear statistic
	$$
	T_{j*}^k = \sum\limits_{i \in \mathcal{B}_{1k}} \mathbf{s}_i,
	$$
	where $\mathcal{B}_{1k}$ here is the first of the two new subgroups 
	that are defined by splitting in split point $k$ of variable $Z_{j*}$. 
	The split point	is then chosen as follows:
	$$ k* = \operatornamewithlimits{argmin}\limits_{k} 
	c(t_{j*}^k, \mu_{j*}^k, \Sigma_{j*}^k).$$
\end{itemize}

\section{Empirical evaluation}\label{app_eval}

\setkeys{Gin}{width=0.95\textwidth}
To check whether the proposed method can recover smooth treatment effect
functions, we evaluated its performance on artificial data. To do so, we
simulated data from a normal linear regression model. We simulated ten correlated
patient characteristics, where only one is in a non-linear interaction with
the treatment. In the following, we compare the log-likelihood of our method to
the log-likelihood of the true underlying model and the naive model that
assumes an overall applicable treatment effect (Section \ref{sec.sim_ll}) and
show the predicted treatment effects in dependence plots (Section
\ref{sec.sim_dp}) and the variable importances of the true predictive
factor and the noise variables (Section \ref{sec.sim_vi}).

We simulated 600 patients, half of which were treated ($x_A = 1$) and half 
of which were untreated
($x_A = 0$). The ten partitioning variables $\mathbf{Z}$ are normally
distributed
\begin{align}
	\mathbf{Z} &\sim \mathcal{N}_{10}(\mathbf{0}, \mathbf{\Sigma}_Z)
\end{align}
and correlated with the covariance matrix
\begin{align*}
	\mathbf{\Sigma}_Z &= \begin{pmatrix} 
		1 & 0.2 & \cdots & 0.2 \\ 
		0.2& 1 & \cdots & 0.2 \\ 
		\vdots & \vdots  & \ddots & \vdots \\
		0.2 & 0.2 & \cdots & 1
	\end{pmatrix}.
\end{align*}
The primary outcome depends on treatment and partitioning variables as follows
\begin{align}
	Y | \mathbf{X}=\mathbf{x}, \mathbf{Z}=\mathbf{z} 
	&\sim \mathcal{N}(1.9 + 0.2 \cdot x_A + 3 \cdot \cos(z_1) \cdot x_A, 1).
\end{align}
In this example the true model parameters are defined as follows:
\begin{align}
	\alpha(\mathbf{z}) &= 1.9\\
	\beta(\mathbf{z}) &= 0.2 + 0.3 \cdot \cos(z_1). \nonumber
\end{align}

\begin{figure}
	\centering
  \includegraphics[width = 0.69\textwidth]{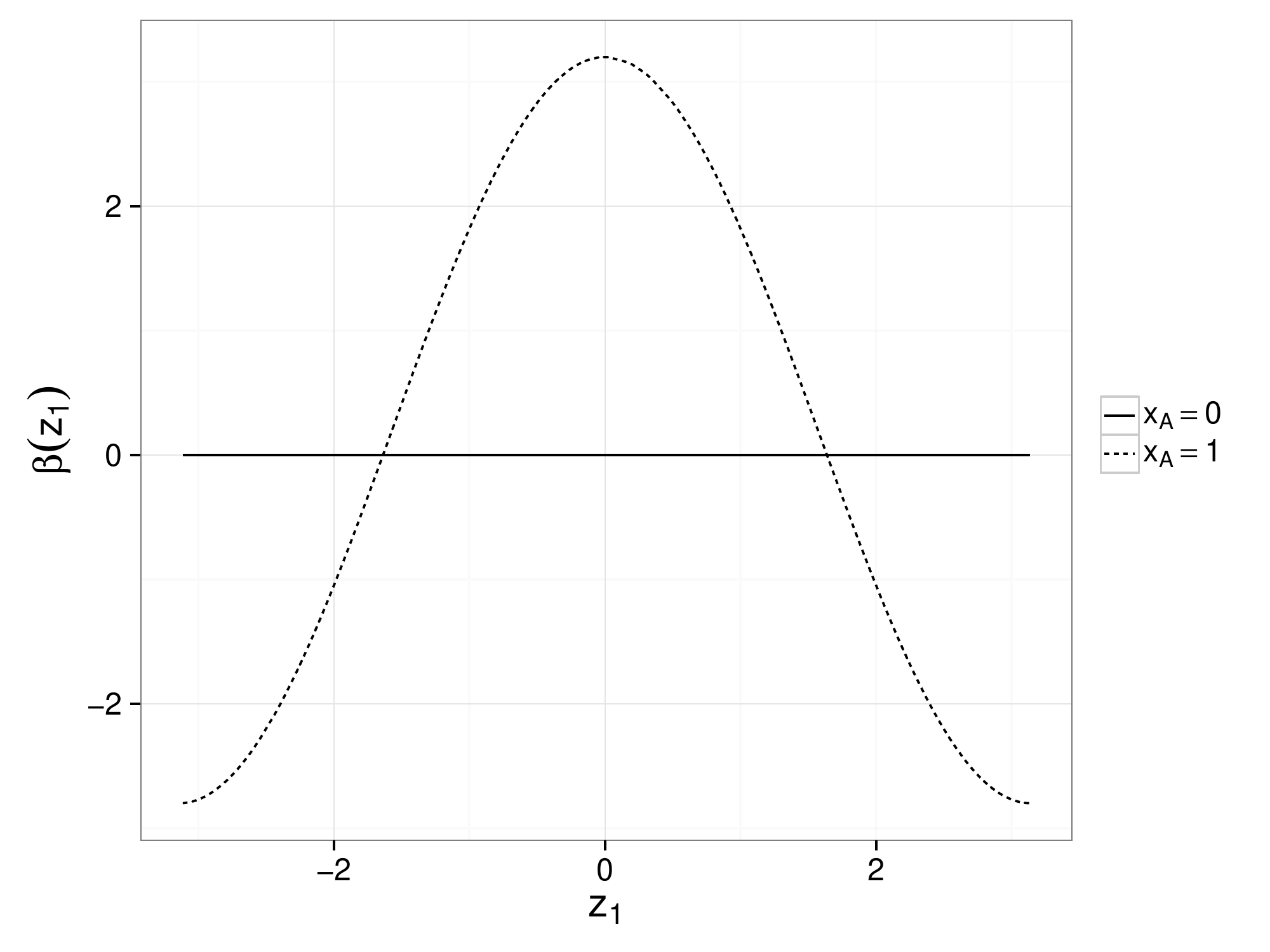}
	\caption{True treatment effect in given simulated data.}
	\label{fig.cos}
\end{figure}

This means that the treatment effect depends on the value of $z_1$ and this
dependency has the form of a cosinus function (see Figure~\ref{fig.cos}).

\subsection{Comparison of models by comparing the log-likelihood}\label{sec.sim_ll}
To compare our method with (i) a correctly specified model taking into account
the main effects of $x_A$ and $\cos(z_1)$ as well as the interaction of $x_A$
and $\cos(z_1)$ and (ii) a simple linear model including only the treatment
$x_A$ as a covariate, we drew 100 learning samples and 100 test samples using
the data simulation procedure explained above and computed the out-of-sample
log-likelihoods (i.e.\ based on the test data) for the models after applying
them to each of the 100 learning data sets. The log-likelihood contributions
\begin{align}
	l\left((y, \mathbf{x})_i, \hat{\boldsymbol{\vartheta}}(\mathbf{z}_i)
	\right) = 
	\left(y_i - \mathbf{x}_i^\top \hat{\boldsymbol{\vartheta}}(\mathbf{z}_i)
	\right)^2
\end{align}
with $\mathbf{x}_i = (1, x_{iA})^\top$ and
$\hat{\boldsymbol{\vartheta}}(\mathbf{z}_i) = (\hat{\alpha}(\mathbf{z}_i),
\hat{\beta}(\mathbf{z}_i))^\top$ are taken from the personalised models of our
method (see Section \ref{sec.vi}). Note that for the simple linear model the
log-likelihood contributions are defined as above, but only with constant
parameters, for the fully specified model $\mathbf{x}_i = (1, x_A, \cos(z_1),
x_A \cdot \cos(z_1))^\top$ and $\mathbf{\hat{\vartheta}} = (\hat{\alpha},
\hat{\beta}_A, \hat{\beta}_{\cos(z_1)}, \hat{\beta}_{A,\cos(z_1)})^\top$.

\begin{figure}
	\centering
  \includegraphics[width = 0.69\textwidth]{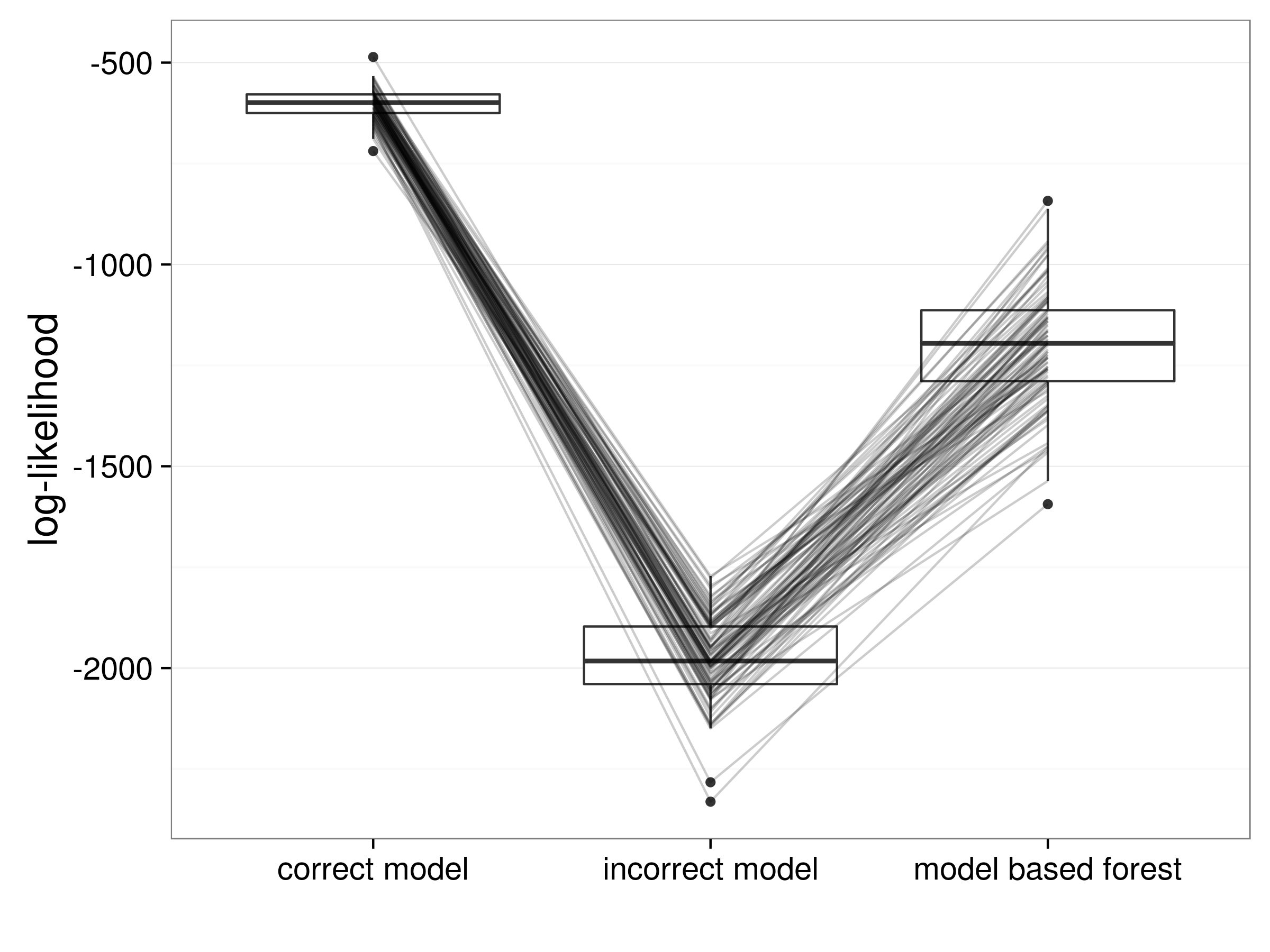}
	\caption{Out-of-sample log-likelihoods obtained from the three models. 
	Each line represents one simulated data set.}
	\label{fig.logliks}
\end{figure}
The log-likelihoods of our method are higher than the log-likelihoods of the
simple and incorrect linear model and lower than the log-likelihoods of the
correctly specified model (Figure~\ref{fig.logliks}). Therefore, we conclude
that our method performs reasonably well.

\subsection{Dependence plots}\label{sec.sim_dp}

\begin{figure}
        \centering
	\begin{subfigure}[t]{0.49\textwidth}	
\includegraphics{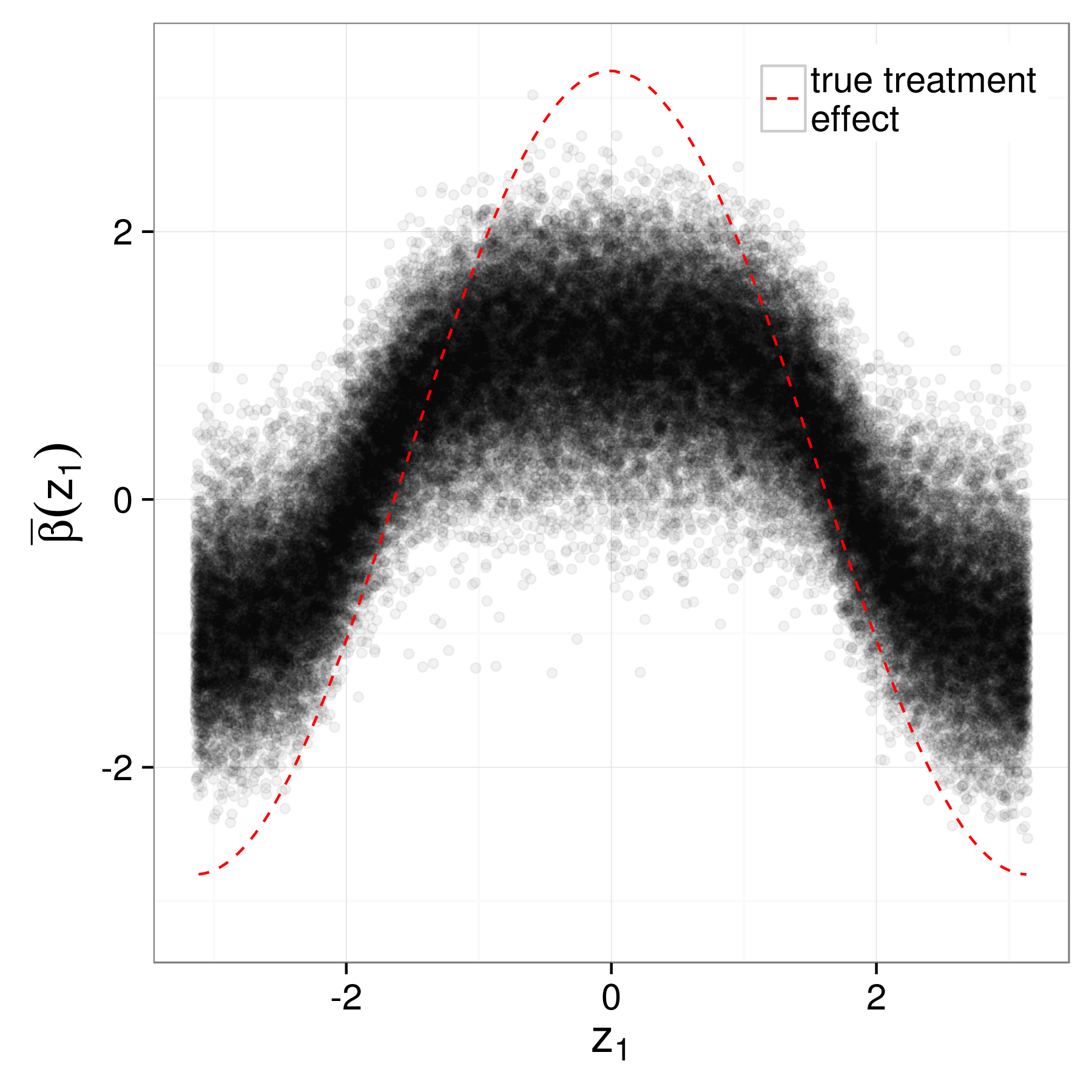}
	        \caption{Dependence plot for $z_1$.}
        	\label{fig.z1}
	\end{subfigure} 
\hfill
        \begin{subfigure}[t]{0.49\textwidth}
\includegraphics{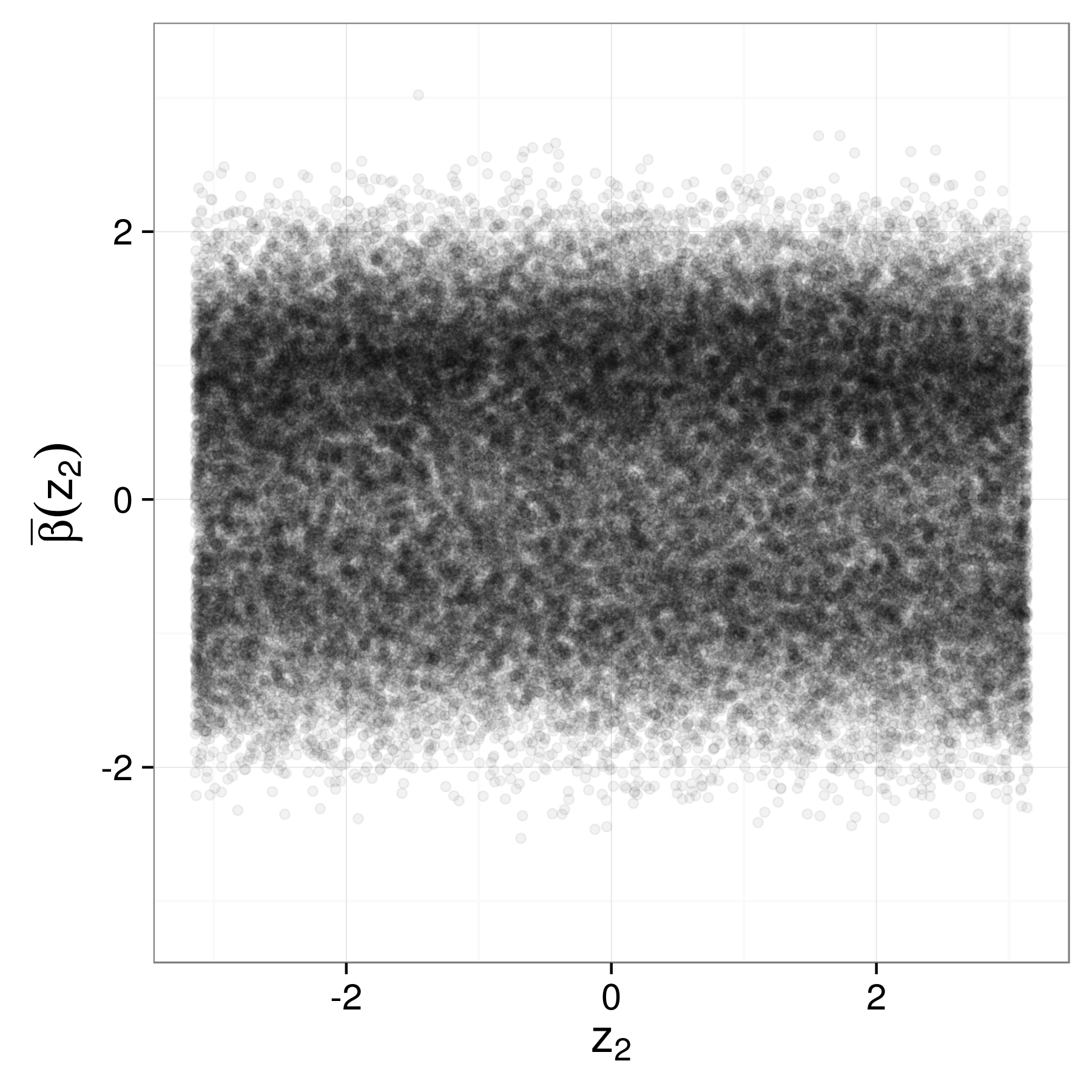}
                \caption{Dependence plot for $z_2$}
                \label{fig.z2}
        \end{subfigure}
\caption{Joint dependence plots of all 100 simulations.}
\label{fig.dp}
\end{figure}
For the same 100 simulated test data sets as above, we obtained the dependence
plots. Figure \ref{fig.dp} shows two dependence plots in which all 100
simulations are combined by layering them on top of each other.  The dependence
plot of partitioning variable $z_1$ (Figure~\ref{fig.z1}) shows a curve that is
fairly similar to that of Figure~\ref{fig.cos}, except that the effect is
shrunken towards zero. Note that with a larger sample or differently tuned
parameters (e.g. larger trees), one could get better results for the extreme
treatment effects.  As expected, for partitioning variables $z_2$ to $z_{10}$,
there is only random fluctuation around zero (see as an example
Figure~\ref{fig.z2}, which shows the dependence plot for $z_2$).

\subsection{Variable importance}\label{sec.sim_vi}
\begin{figure}[htp]
        \centering
        \includegraphics[width=0.6\textwidth]{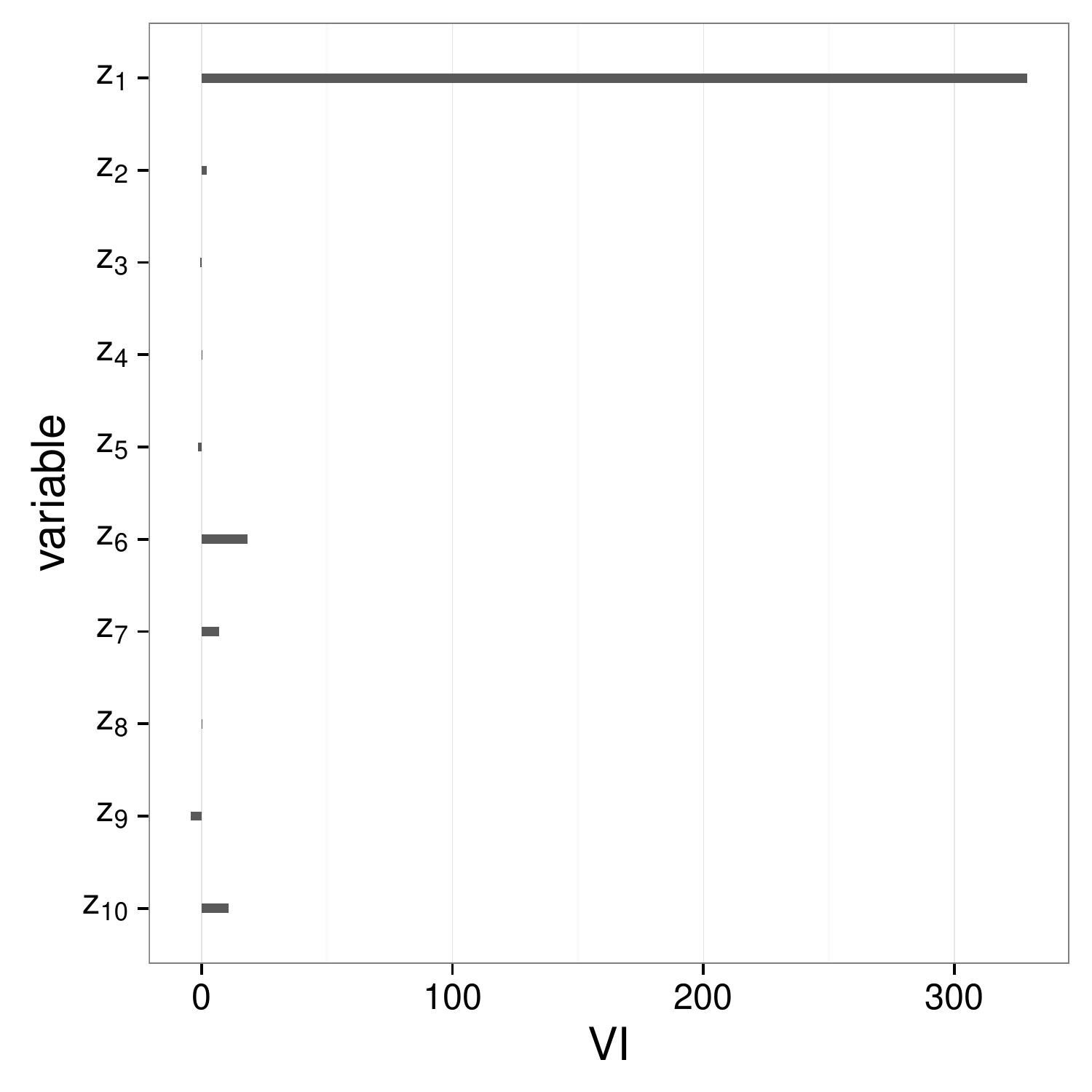}
        \caption{Variable importances of the predictive factor $z_1$ and 
	noise variables $z_2$ to $z_{10}$.}
        \label{fig.vari}
\end{figure}
Variable importances for one simulated data set are shown in
Figure~\ref{fig.vari}. As expected, partitioning variable $z_1$ is the only
variable with a clearly positive variable importance. Even though all
partitioning variables are correlated, the method was able to distinguish
between the correlation and predictive effect.

\end{appendix}

\end{document}